\documentclass[trackchanges, twocolumn]{aastex701}

\usepackage{booktabs}
\usepackage{xcolor}
\usepackage{amsmath,siunitx}
\usepackage{multirow}

\shorttitle{Multimodality for real-time classification of transients}
\shortauthors{Shah et al.}

\newcommand{\NU}{
Department of Physics and Astronomy, Northwestern University, 2145 Sheridan Road, Evanston, IL 60208, USA}
\newcommand{\CIERA}{
Center for Interdisciplinary Exploration and Research in Astrophysics (CIERA), 1800 Sherman Ave, Evanston, IL 60201, USA}
\newcommand{\SkAI}{
NSF-Simons AI Institute for the Sky (SkAI), 172 E. Chestnut St., Chicago, IL 60611, USA}

\newcommand{\lite}{\texttt{ORACLE-2 Lite}}
\newcommand{\base}{\texttt{ORACLE-2}}
\newcommand{\omni}{\texttt{ORACLE-2 Omni}}

\definecolor{Tomato}{HTML}{FF6347}

\definecolor{SteelBlue}{HTML}{4682B4}

\definecolor{DarkMagenta}{HTML}{8B008B}

\definecolor{red}{HTML}{FF0000}

\begin{document}

\title{Leveraging Multimodality for Real-Time Classification of Transients and Variables found by the Zwicky Transient Facility}

\correspondingauthor{Ved~G.~Shah}
\email{vedshah2029@u.northwestern.edu}

\author[orcid=0009-0009-1590-2318]{Ved~G.~Shah}
\affiliation{\NU}
\affiliation{\CIERA}
\affiliation{\SkAI}
\email[]{}  

\author[0000-0002-5683-2389]{Nabeel~Rehemtulla}
\affiliation{\NU}
\affiliation{\CIERA}
\affiliation{\SkAI}
\email[]{}  

\author[0000-0001-9515-478X]{Adam~A.~Miller}
\affiliation{\NU}
\affiliation{\CIERA}
\affiliation{\SkAI}
\email[]{}  

\author[0000-0003-1314-4241]{Sushant~Sharma~Chaudhary}
\affiliation{School of Physics and Astronomy, University of Minnesota, Minneapolis, Minnesota 55455, USA}
\email[]{}  

\author[0000-0002-8262-2924]{Michael~W.~Coughlin}
\affiliation{School of Physics and Astronomy, University of Minnesota, Minneapolis, Minnesota 55455, USA}
\email[]{}  

\author[0009-0009-7000-8343]{Antoine~Le~Calloch}
\affiliation{School of Physics and Astronomy, University of Minnesota, Minneapolis, Minnesota 55455, USA}
\email[]{}  

\author[0000-0002-3168-0139]{Matthew~J.~Graham}
\affiliation{Division of Physics, Mathematics, and Astronomy, California Institute of Technology, Pasadena, CA 91125, USA}
\email[]{}  

\author[0000-0002-0987-3372]{Joahan~Castaneda~Jaimes}
\affiliation{Division of Physics, Mathematics, and Astronomy, California Institute of Technology, Pasadena, CA 91125, USA}
\email[]{}  

\author[0009-0003-6181-4526]{Theophile~Jegou~du~Laz}
\affiliation{Division of Physics, Mathematics, and Astronomy, California Institute of Technology, Pasadena, CA 91125, USA}
\email[]{}  

\author[0000-0003-2242-0244]{Ashish~A.~Mahabal}
\affiliation{Division of Physics, Mathematics, and Astronomy, California Institute of Technology, Pasadena, CA 91125, USA}
\affiliation{Center for Data Driven Discovery, California Institute of Technology, Pasadena, CA 91125, USA}
\email[]{}  

\author[0000-0002-8532-9395]{Frank~J.~Masci}
\affiliation{IPAC, California Institute of Technology, 1200 E. California Blvd, Pasadena, CA 91125, USA}
\email[]{}  

\author[0000-0003-1227-3738]{Josiah~Purdum}
\affiliation{Caltech Optical Observatories, California Institute of Technology, Pasadena, CA 91125, USA}
\email[]{}  

\author[0000-0002-0387-370X]{Reed~Riddle}
\affiliation{Caltech Optical Observatories, California Institute of Technology, Pasadena, CA 91125, USA}
\email[]{}  

\author[0000-0003-1546-6615]{Jesper~Sollerman}
\affiliation{The Oskar Klein Centre, Department of Astronomy,
AlbaNova, SE-106 91 Stockholm , Sweden}
\email[]{}  

\author[0009-0002-1319-3975]{Anastasia~Wei}  
\affiliation{Department of Astronomy, University of California, University Drive, Berkeley, CA 94720, USA}
\email[]{} 

\author[0000-0002-5619-4938]{Mansi~M.~Kasliwal}
\affiliation{Division of Physics, Mathematics, and Astronomy, California Institute of Technology, Pasadena, CA 91125, USA}
\email[]{}

\begin{abstract}

Modern time-domain surveys such as the Zwicky Transient Facility (ZTF) generate hundreds of thousands of alerts each night, making real-time decisions for follow-up observations a central challenge in time-domain astronomy. Robust early classification is crucial for making informed decisions, but is hindered by sparse light curves and degeneracies between classes. In this work, we leverage multimodality to substantially improve real-time classification and demonstrate the practicality of our approach by deploying our model on the ZTF alert stream. Building on the Online Ranked Astrophysical CLass Estimator (\texttt{ORACLE}), we introduce the \texttt{ORACLE-2} models, which combine light curves, metadata, and images for real-time hierarchical classification. Using both real and simulated datasets, we show that incorporating additional modalities consistently improves classification performance. On observations from ZTF's Bright Transient Survey, our best-performing model, \texttt{ORACLE-2 Omni}, achieves a macro F1 score of 0.73 -- an improvement of up to 11\% over models using light curves and metadata alone, and up to 40\% over light-curve-only models, with the strongest gains realized at early times. To demonstrate applicability to the Legacy Survey of Space and Time, which will increase alert volume by more than an order of magnitude, we train a light curve + metadata variant on the simulated \texttt{ELAsTiCC} dataset. This model achieves a macro F1 score of 0.88, an improvement of up to 13\% over the light-curve-only variant, matching the performance of other state-of-the-art models. Finally, we quantify the trade-offs between performance and throughput, identifying regimes where multimodal approaches offer the greatest benefit. These results show that combining multiple modalities improves early-time classification, enabling more effective triage of high-volume alert streams for current and future time-domain surveys.

\end{abstract}


\section{Introduction}
\label{sec:introduction}
The latest generation of time-domain surveys such as the Vera C. Rubin Observatory’s Legacy Survey of Space and Time \citep[LSST;][]{2019Ivezic_LSST} and the Argus Optical Array \citep{Law_2022_argus} will generate an unprecedented volume of data, reaching millions of alerts\footnote{An alert is an automated notification issued when a telescope detects a statistically significant variation in flux.} per night. This data deluge makes real-time decision-making a core challenge, which, in turn, involves determining which sources merit additional follow-up. Because spectroscopic resources are expected to be over-subscribed by several orders of magnitude \citep{Kulkarni20, 4most_tides}, there is a growing need for automated systems that can reliably classify sources and surface scientifically valuable ones based on survey photometry alone.

Machine learning has increasingly been adopted to address this bottleneck in operational settings. For example, the Dark Energy Survey used ML-based classifiers to identify Type Ia supernovae for cosmological analyses \citep{descollaboration2025}, while the Zwicky Transient Facility \citep[ZTF;][]{ZTF1, ZTF2, ZTF4_Masci19, ZTF3_Dekany2020} employs models for automated spectroscopic follow-up of (young) transients \citep{BTSBot1, BTSbot-nearby}, identifying Type Ia supernovae \citep{Fremling21_SNIascore}, identifying core-collapse supernova \citep{CCScore_25}, general photometric classification \citep{apple_cider_25}, etc. These systems demonstrate the feasibility of data-driven prioritization.

Effective prioritization of sources requires, among other things, reliable classifications; yet early observations typically lack information about the light curve morphology, making fine-grained classification difficult. Thus, acting early on limited information risks allocating precious spectroscopic resources to commonplace sources, while waiting for a confident photometric classification risks missing temporally evolving emission from the very sources worth observing. In this regime, it is worth building systems that can produce classifications at different levels of details based on the amount of available information. Hierarchical classification provides one such framework by organizing classes into a taxonomy that progresses from coarse to increasingly granular categories. Thus, we can use labels at the top of the taxonomy at early times for confident, albeit less detailed, classifications while additional observations enable more detailed classifications at lower levels in the taxonomy.

The \textbf{O}nline \textbf{R}anked \textbf{A}strophysical \textbf{CL}ass \textbf{E}stimator \citep[henceforth \texttt{ORACLE};][]{oracle}, introduced real-time, hierarchical classification for LSST, allowing for a single model to perform classification along a taxonomy. Unlike conventional approaches, \texttt{ORACLE} produces classification at several levels of detail (Transient vs Variable, Supernova vs Fast Transient vs Long Transient vs Periodic Variable Stars, etc) allowing for useful outputs to be produced from the earliest alerts.

This formulation naturally benefits from the different information contained in the available data modalities. For instance, images provide valuable context about the environment of the source, enabling reliable classification at the top of the hierarchy at early times. Then, as the model progressively ingests the temporally evolving data from the photometric observations, it can produce reliable granular classifications that are not accessible from the image modality alone. In this way, different modalities play complementary roles across the source's temporal evolution. 

In this work, we expand the \texttt{ORACLE} family of models by introducing five new multimodal hierarchical classifiers and demonstrate their efficacy via deployment on the ZTF alert stream.\footnote{Throughout this paper, the term \textit{classification} refers to photometric classification unless otherwise stated. Whenever spectroscopic labels are used, we explicitly state this fact.} These new (\base) models are capable of classifying sources from ZTF and LSST in real time using a combination of light curves, metadata, and images, allowing for the triaging and follow-up of scientifically valuable sources. 

This paper is structured as follows. Section~\ref{sec:dataset} describes the datasets used to develop the models presented in this work. Sections~\ref{sec:arch} and \ref{sec:training} detail the model architectures and training procedures, respectively. Section \ref{sec:results} presents the classification performance of the models, their throughput in real-time applications, and common failure modes. We discuss real-time deployment and related problems in Section~\ref{sec:results:deployment} and \ref{sec:discussion}. Finally, Section \ref{sec:conclusion} summarizes the key conclusions and takeaways, followed by acknowledgments in Section~\ref{sec:acknowledgements}.

\section{Datasets and Taxonomies}
\label{sec:dataset}
\begin{figure}
    \centering
    \includegraphics[width=0.45\linewidth]{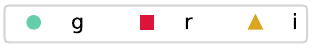}
    \includegraphics[width=0.905\linewidth]{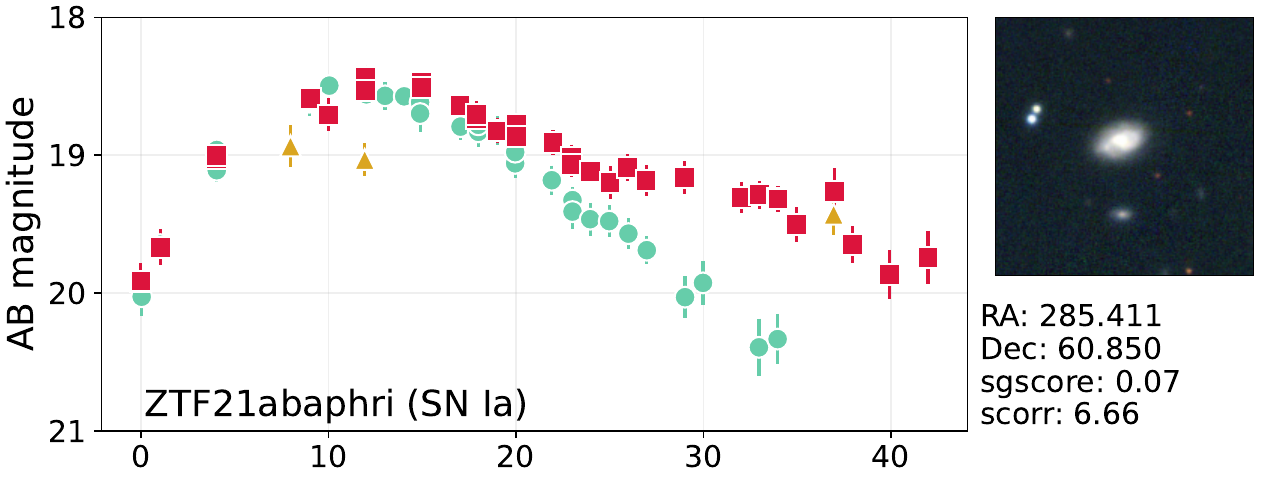}
    \includegraphics[width=0.905\linewidth]{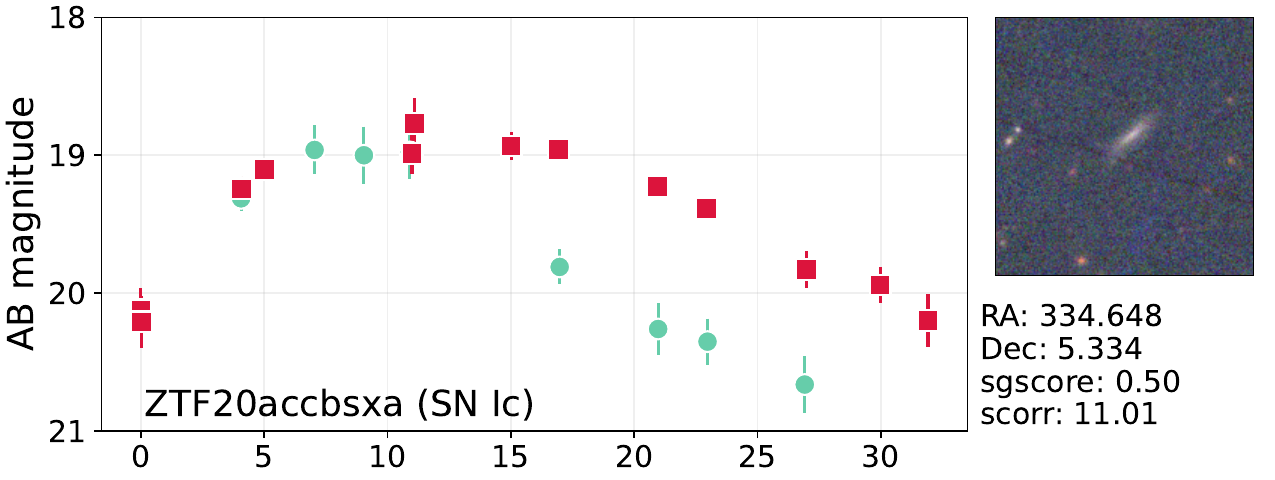}
    \includegraphics[width=0.905\linewidth]{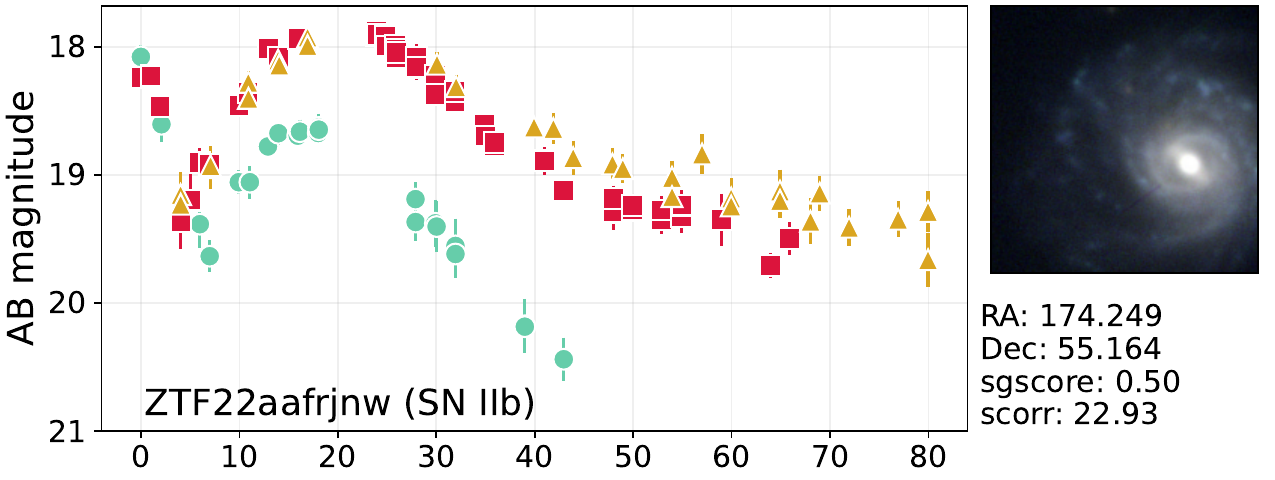}
    \includegraphics[width=0.905\linewidth]{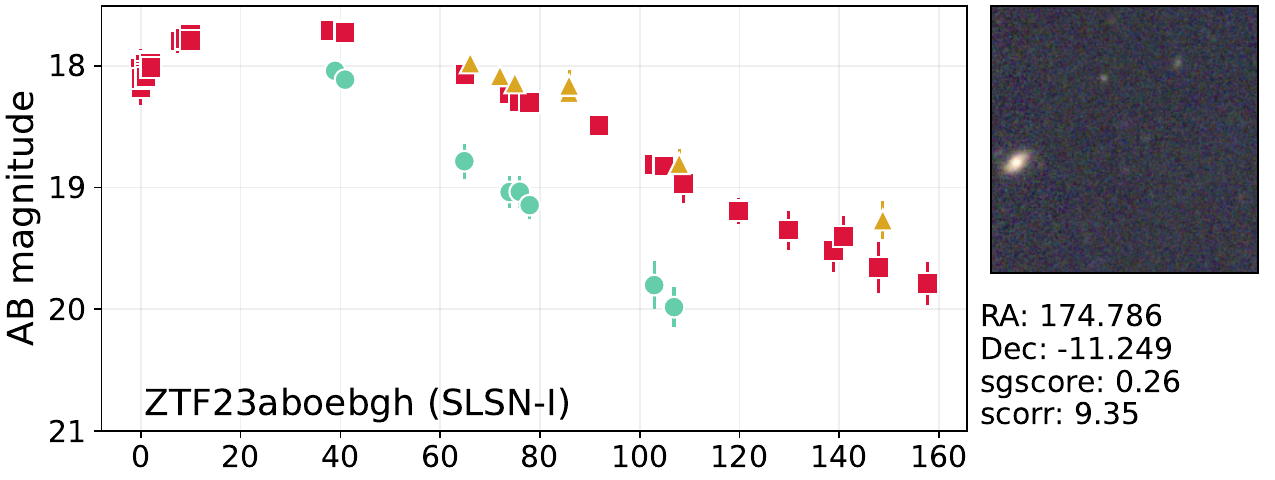}
    \includegraphics[width=0.905\linewidth]{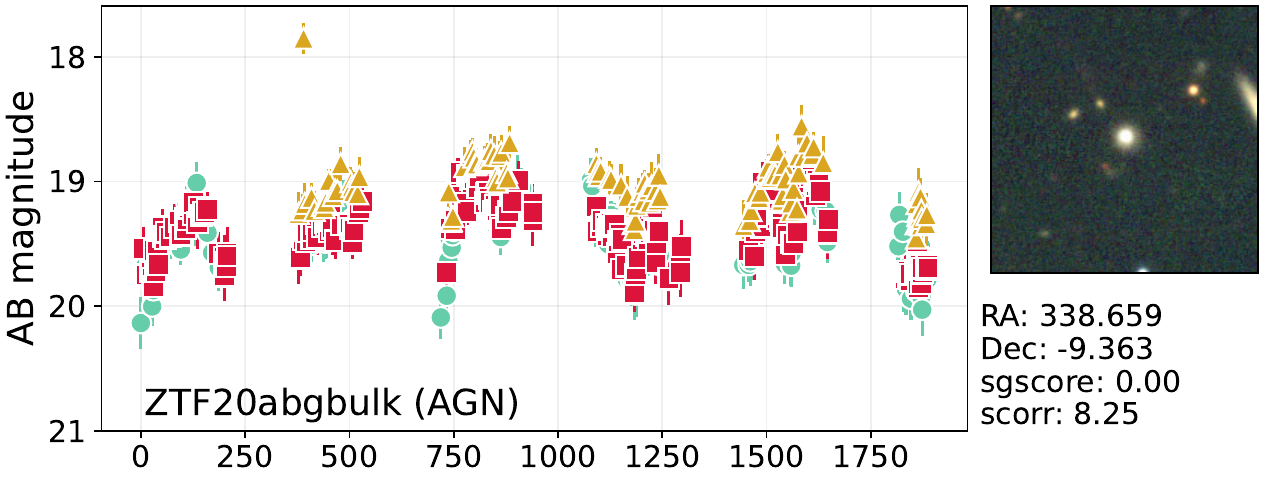}
    \includegraphics[width=0.905\linewidth]{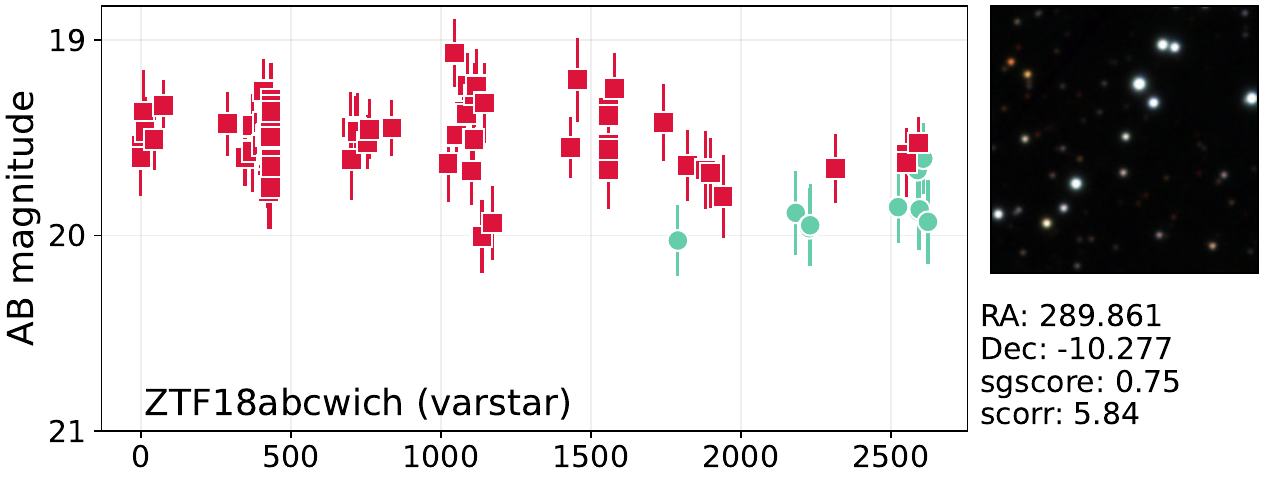}
    \includegraphics[width=0.905\linewidth]{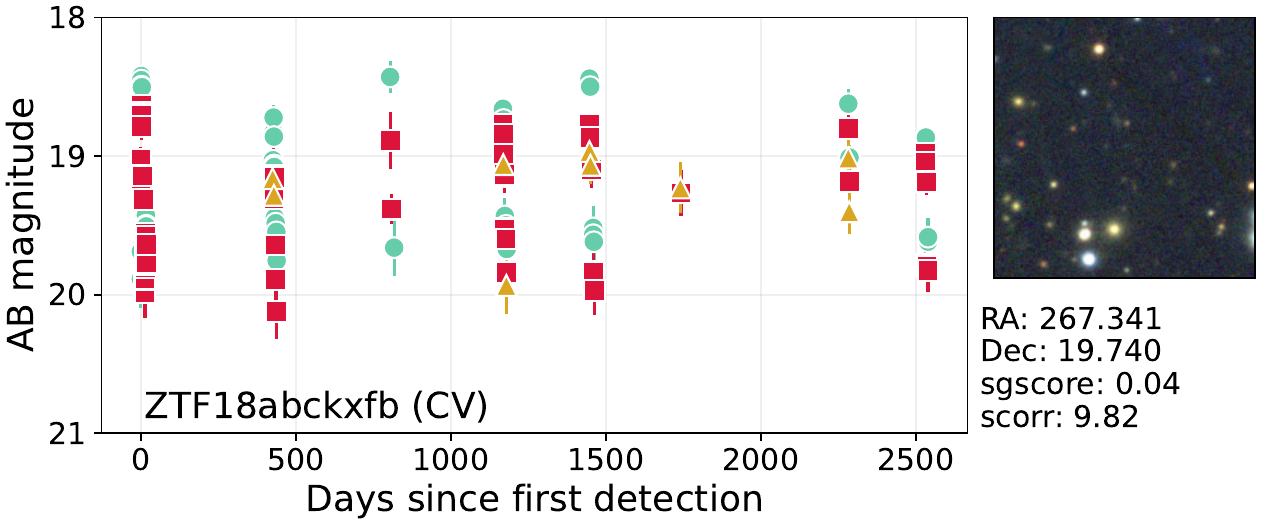}
    \caption{Example light curves and metadata from the BTS dataset, along with the $g$, $r$, and $i$ band images from PanSTARRS-1 for all seven leaf classes in our taxonomy.}
    \label{fig:lc_examples}
\end{figure}

In order to train the models described in this work, we make use of both real (Section \ref{sec:dataset:BTS}) and simulated datasets (Section \ref{sec:dataset:ELAsTiCC}) for ZTF and LSST, respectively. In this section, we discuss the datasets, the taxonomies, and our motivations for including the features used in training.

\subsection{ZTF (Bright Transient Survey)}
\label{sec:dataset:BTS}

The Bright Transient Survey \citep[BTS; ][]{BTS1, BTS2, BTSBot1}, one of the founding science cases of ZTF \citep{ZTF1}, is a magnitude-limited spectroscopically complete extragalactic transient survey utilizing ZTF data. BTS ran from April 2018 through the end of 2024 and produced a large ($>$10,000 sources), highly complete ($\sim\,95$\% completeness) spectroscopic sample of extragalactic transients with $m_\textrm{peak}\lesssim18.5$ mag. The extragalactic transients compiled by BTS alongside several thousand cataloged Active Galactic Nuclei (AGN) and Cataclysmic Variables (CVs) serve as the basis for our training set. Our final dataset contains spectroscopically labeled supernovae (SNe) across several sub-classes including Type Ia supernovae (SNe\,Ia), Type Ib/c supernovae (SNe\,Ib/c), Type II supernovae (SNe\,II), and Type I superluminous supernovae (SLSNe\,I) \citep[see][for details on each class]{Gal-Yam17_handbook}; as well as common persistent sources such as AGN, CVs, and Variable Stars.

Fritz, the ZTF collaboration's instance of SkyPortal \citep{vanderWalt19_skyportal, Coughlin23_skyportal}, was used for vetting bright transient candidates in BTS. Sources cataloged by BTS are made available on Fritz and/or on the BTS Sample Explorer\footnote{\url{https://sites.astro.caltech.edu/ztf/bts/explorer.php}} \citep{BTS2}. We curate our training set with queries to these platforms, as detailed individually for each class below.

\textbf{AGN:} AGN are luminous, compact regions at the centers of galaxies, powered by gas and dust accreting onto a central supermassive black hole. They exhibit stochastic photometric variability over a wide range of timescales that can be difficult to distinguish from supernovae when only sparse or early-time observations are available. See \cite{AGN_review_Padovani_2017} for a review on AGN. To find AGNs for our dataset, we queried both \texttt{Fritz} and the BTS Sample Explorer and selected all sources classified as AGN, CLAGN, QSO, NLS1, Blazars, or BL Lac. Our dataset contains a total of 3,414 of these sources grouped under the AGN class. 

\textbf{CVs:} CVs are white dwarfs in binary systems that exhibit irregular increases in brightness due to mass transfer from their companion. Many CVs exhibit photometric outbursts that can resemble supernovae in sparse or early-time light curves. See \cite{novae_review_Della_Valle2020} for a review on CVs and \cite{Paula21_CV} for examples of CVs observed by ZTF. Similarly to the AGN, we queried both \texttt{Fritz} and the BTS Sample Explorer and selected all sources classified as CV, AMCVn, or novae and group them under the CV umbrella. Our dataset contains a total of 1,109 CVs. 

\textbf{Varstars:} Variable stars constitute another major source of contamination in searches for extragalactic transients. Given the breadth and diversity of this population, a comprehensive discussion of variable star subclasses is beyond the scope of this work. Instead, our objective is to classify variable stars that enter the BTS alert stream despite the filtering criteria designed to select transient sources. To construct this sample, we cross-match BTS candidates with the \cite{ztf_var_catalog} ZTF variable star catalog, searching for matches within $\ang{;;2}$ of each BTS candidate. This ensures that we only include variable stars that contaminate the BTS sample rather than the broader variable-star population. Our dataset contains a total of 769 variable stars.

\textbf{SNe\,Ia:} Type Ia supernovae are thermonuclear explosions which take place when a white dwarf in a binary system approaches the Chandrasekhar limit, leading to the complete disruption of the star. Unlike core-collapse SNe, SNe\,Ia are also known to occur in older, redder, and elliptical galaxies \citep{Foley_Mandel13_host_galaxy}. See \cite{Ia_review_Howell_2011} for a review on SNe\,Ia and \cite{Rigault25_SNIa_DR2} for examples of SNe\,Ia observed by ZTF. We collect our SN\,Ia sample by grouping any sources classified as SN\,Ia, SN\, Ia-00cx, SN\,Ia-03fg, SN\,Ia-91T, SN\,Ia-91bg, SN\,Ia-99aa, SN\,Ia-CSM, SN\,Ia-pec, or SN\,Iax under the SN\,Ia umbrella. Our dataset contains a total of 6,871 SNe\,Ia. 

\textbf{SNe\,Ib/c:} Type Ib/c supernovae are hydrogen-poor SNe resulting from the core-collapse of stars which had their outer layer(s) of hydrogen (and helium, in the case of Type Ic) stripped as a result of stellar winds or via mass transfer due to interactions with a binary companion. Since their progenitors are massive stars, they preferentially occur in star-forming regions of spiral or irregular galaxies \citep{Kelly12_CC_Host, Hakobyan14_SpiralHosts}. See \cite{SESNe_properties_Prentice19, Sollerman22_Ibc_ZTF} for an overview of SNe\,Ib/c properties. We collect our SNe\,Ib/c sample by grouping any sources classified as SN\,Ib, SN\,Ib-pec, SN\,Ib/c, SN\,Ibn, SN\,Ic, SN\,Ic-BL, SN\,Ic-SL, or SN\,Icn under the SNe\,Ib/c umbrella. Our dataset contains a total of 521 SNe\,Ib/c. 

\textbf{SNe\,II:} Type II supernovae are hydrogen-rich core-collapse SNe. Like other core-collapse SNe (CCSNe), SNe\,II also have massive-star progenitors and preferentially occur in star-forming regions of younger galaxies \citep{Kelly12_CC_Host}. See \cite{Woosley05_CCSNe_review} for an overview of CCSNe, including SN II \citep{Hinds25_SNeII_ZTF}. We collect our SNe\,II sample by grouping any sources classified as SLSN\,II, SN II, SN\,II-SL, SN\,II-pec, SN\,IIL, SN\,IIP, SN\,IIb, SN\,IIb-pec, or SN\,IIn under the SNe\,II umbrella. Our dataset contains a total of 1918 SNe\,II. 

\textbf{SLSNe\,I:} Type-I superluminous supernovae are extremely bright core-collapse explosions with luminosities $\gtrsim10$ times higher than the typical SNe, and are thought to originate from particularly massive stars. There are several theorized models for explaining SLSNe\,I \citep[see Section 1 of][and references within]{Gomez24_SLSNeICatalog}, although no consensus has emerged. There is also evidence that the host galaxies of SLSNe\,I are different from the hosts of more ``typical" CCSNe, with these SNe generally occurring in metal-poor dwarf galaxies \citep{Neill11,Lunan15_SLSNe_environments,Perley16,Schulze21}. See \cite{SLSN_review_Gal_Yam_2019} for a review on SLSNe\,I and \cite{Chen23_SLSNe_ZTF} for examples of SLSNe observed by ZTF. We collect our SLSNe\,I sample by grouping any sources classified as SLSN\,I or SLSN\,I.5 under the SLSNe\,I umbrella. Our dataset contains a total of 91 SLSNe\,I. 

While constructing this dataset, we made the very intentional choice to leave in common contaminants (such as AGN, CVs, and variable stars) and peculiar subtypes (such as SN\,Ia CSM, SN\,Iax, SN\,Icn etc.). This was done in order to build a single system that can classify typical SNe, peculiar SN subtypes, \textit{and} identify common contaminants. Additionally, we do \textit{not} apply any quality cuts to our data based on the number of detections, signal to noise, presence of multiple bands, etc., except for the minimal filtering done by the BTS alert filter \citep{BTS2}. This choice is especially important while reporting real-time performance metrics, since excluding either peculiar subtypes or contaminants can lead to misleading or overconfident statistics. Other rare classes of objects such as tidal disruption events, fast blue optical transients, and $\gamma$-ray burst afterglows, make up a very small fraction of our dataset ($<0.8\%$) but do not fit into our taxonomy. We leave them out for the time being and discuss their impact in Section \ref{sec:results:failures}.

We make use of the photometry, metadata, and reference images included in the ZTF alert packets (see Table \ref{table:bts_featues}) for every source in our sample. Although BTS only uses ZTF public survey data, we include ZTF partnership survey data in our training set to increase the overall size and quality of our data. This also involves adding $i_{{\text{ZTF}}}$-band data, which were absent from the ZTF public stream during BTS. The data used in this work were downloaded from Fritz \citep{vanderWalt19_skyportal, Coughlin23_skyportal} and the legacy internal ZTF alert broker Kowalski \citep{kowalski_braai}. While the BTS data constitutes a large fraction of our training data, we also include observations and labels obtained after the official conclusion of the survey.
% All labels for this dataset except for variable stars, as discussed above, are obtained from the \texttt{BTS Sample Explorer} \citep[][]{BTS2}.
Our final dataset contains 14,798 sources and has a cutoff date of December 29, 2025. We use a $80\%/10\%/10\%$ stratified train/validation/test split for this dataset, ensuring that representative class imbalance is maintained across all three splits. 

\subsubsection{Additional Features and Modalities}

\begin{figure}
    \centering
    \includegraphics[width=\linewidth]{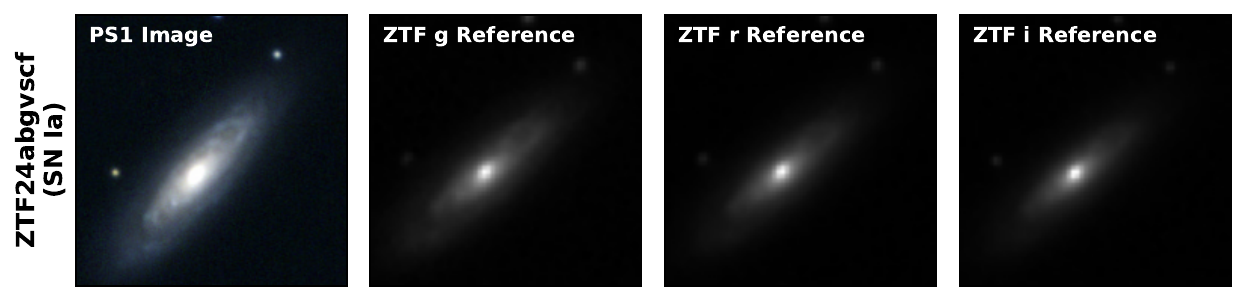}
    \includegraphics[width=\linewidth]{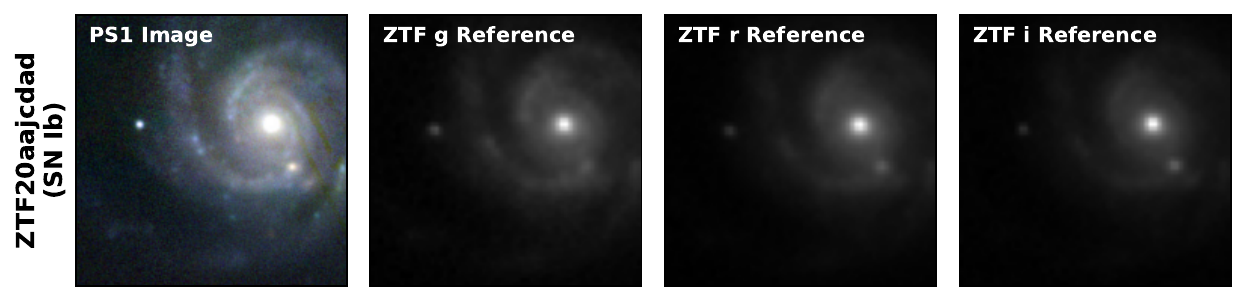}
    \includegraphics[width=\linewidth]{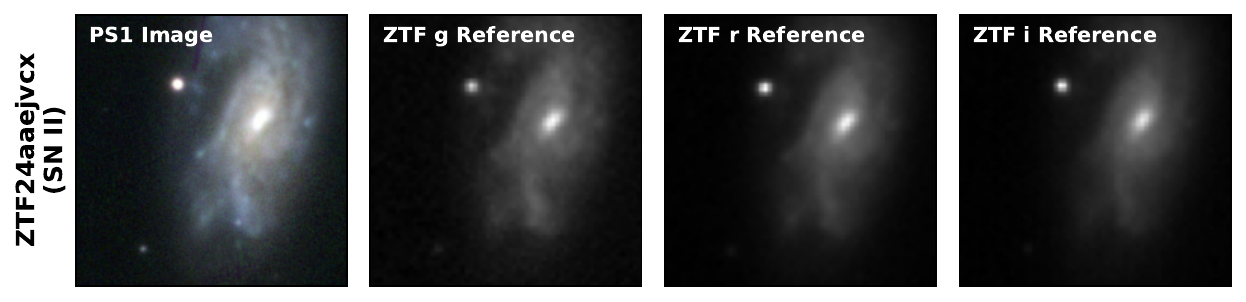}
    \includegraphics[width=\linewidth]{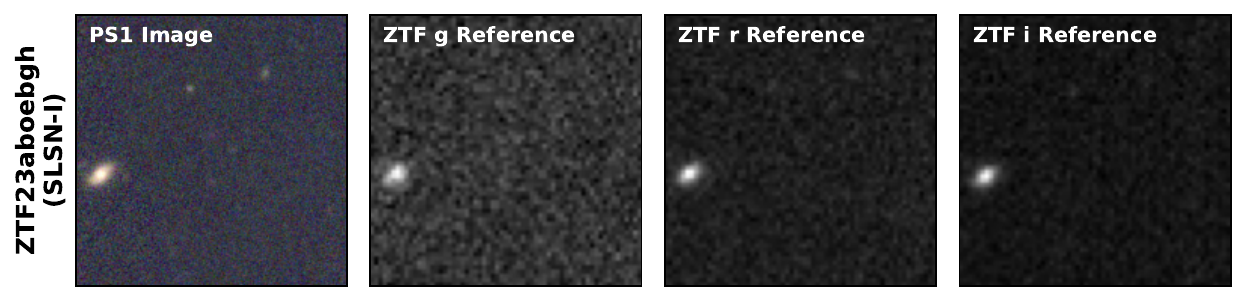}
    \includegraphics[width=\linewidth]{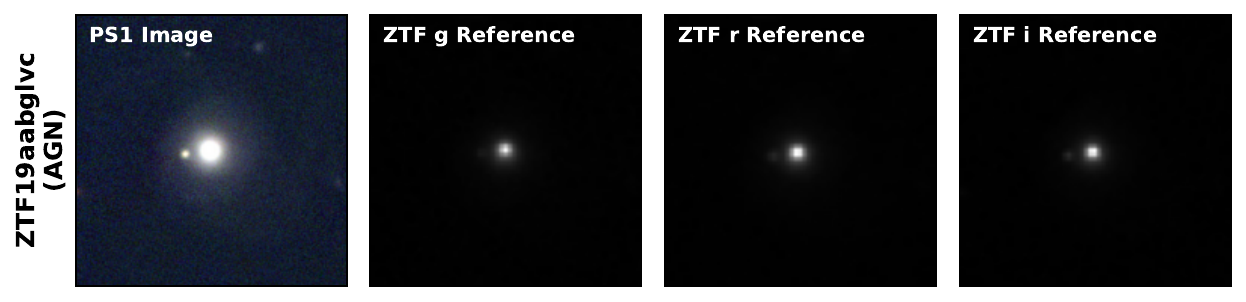}
    \includegraphics[width=\linewidth]{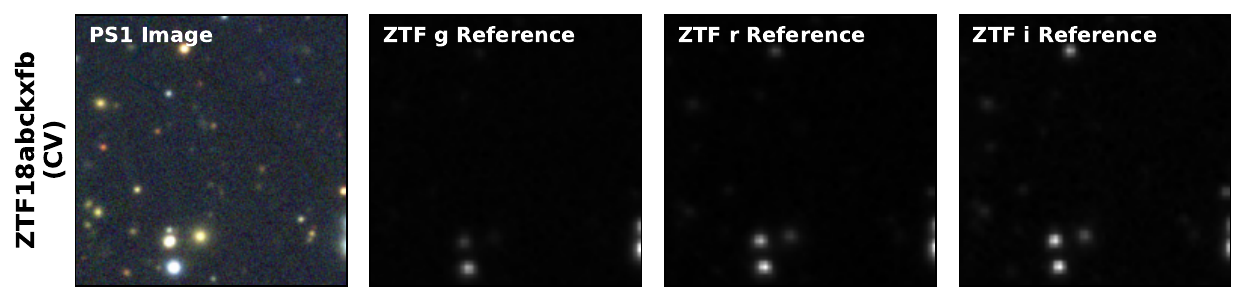}
    \includegraphics[width=\linewidth]{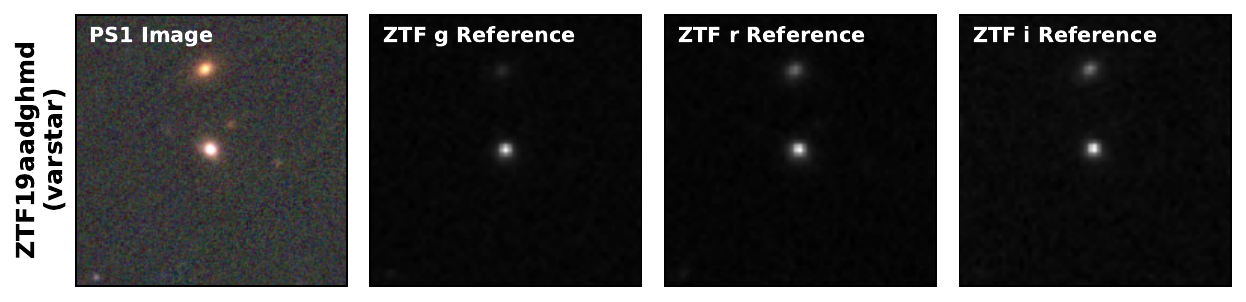}
    \caption{Representative images for different classes in the BTS dataset. Each row corresponds to a different object, while columns (left to right) show the PS1 $gri$ image, the ZTF $g$-band, the ZTF $r$-band, and the ZTF $i$-band reference images.}
    \label{fig:image_comparisons}
\end{figure}

In addition to the light curves and metadata from the alerts, we augment our dataset with additional modalities and contextual information. Specifically, we cross-match every source in our sample with the \texttt{AllWISE} catalog \citep{wise_catalog}, to find matches within $\ang{;;2.75}$. We then collate the \textit{W1}, \textit{W2}, \textit{W3}, and \textit{W4} magnitudes for the closest match, which are used as features to help isolate AGNs. This choice is motivated by the relative isolation of AGN in the $(W1 - W3)$ vs $(W2 - W3)$ color space \citep[e.g.,][]{Assef_AGN_WISE}. 

Next, we add images from two different surveys to our dataset. This decision is motivated by the fact that persistent sources, especially ones with galactic origins, such as CVs and Variable Stars, often occur in crowded fields close to the Galactic plane and thus should be easy to identify in images (see Figure \ref{fig:lc_examples} for examples), aiding in early time classification. Additionally, several studies have shown that SN classes are correlated with the properties of the host galaxies that they occur in \citep{Neill11, Kelly12_CC_Host, Foley_Mandel13_host_galaxy, Hakobyan14_SpiralHosts, Lunan15_SLSNe_environments, Perley16, ghost,Schulze21}, including their morphology and color. Thus, the addition of images should provide rich, contextual information to our models, which could enable better classification. It is worth noting that while images are used as input for training some of our models, we do not perform explicit host galaxy association as in \cite{ghost} or \cite{Villar25_Hierarchical_Classification}. Thus, any host galaxy associations and features are implicitly learned by the network as part of the training process. As shown in Section \ref{sec:results}, this approach proves effective, with images playing a particularly important role in improving early-time performance for real-time classification. In this work, we experiment with two different sources for our images: 

\textbf{1. Triple Channel Pan-STARRS-1 Images:} These are obtained from the Panoramic Survey Telescope and Rapid Response System \citep[Pan-STARRS-1 or PS1;][]{Kaiser02_PS, Chambers16_PS} by querying the sky location of the first alert for each source and producing a $252\times252\times 3$ $(\text{height}\times\text{width}\times \text{channels})$ pixels cutout with the $g_{\text{PS1}}$, $r_{\text{PS1}}$, and $i_{\text{PS1}}$ band images. For a small fraction ($<0.02\%$) of our dataset, one or more of the $g$, $r$, or $i$ channels were missing. In these instances, we zero out the missing channel while producing the cutout with the available channels to maintain the dimensionality of the input. These images have a pixel scale of 0.25"/pixel, resulting in a 63"$\times$63" image. This field-of-view is chosen to match that of the ZTF cutouts discussed next. Examples of light curves from all 7 classes, along with the corresponding PS1 $gri$ images and selected metadata, are shown in Figure \ref{fig:lc_examples}. While the PS1 images are high-quality and multi-channel, most brokers do not have the infrastructure to host a cutout service, making the real-time deployment of models utilizing this data non-trivial. This motivates our second choice.

\textbf{2. Single Channel ZTF Reference Images:}  We also test variants of our multimodal models with the ZTF images included in the alerts. This represents the more practical deployment scenario since it exclusively uses image data contained within the alert packets, thus removing reliance on any external dependencies. Instead of including the science, reference, and difference cutouts provided individually in each alert, we focus on the reference images and create a $63\times63\times 3$ $(\text{height}\times\text{width}\times \text{channels})$ tensor of ZTF $gri$ reference images. The image fed to the model only passes the channel corresponding to the latest alert's filter and zeros out the other channels. These images have a pixel scale of 1"/pixel, resulting in a 63"$\times$63" image. 

Figure \ref{fig:image_comparisons} highlight the difference between triple channel Pan-STARRS-1 images and the single channel ZTF reference images. A complete list of the features used to train the BTS models, along with brief descriptions, is provided in Table \ref{table:bts_featues}.

\subsubsection{Choice of taxonomy}

\begin{figure}
    \centering
    \includegraphics[width=\linewidth]{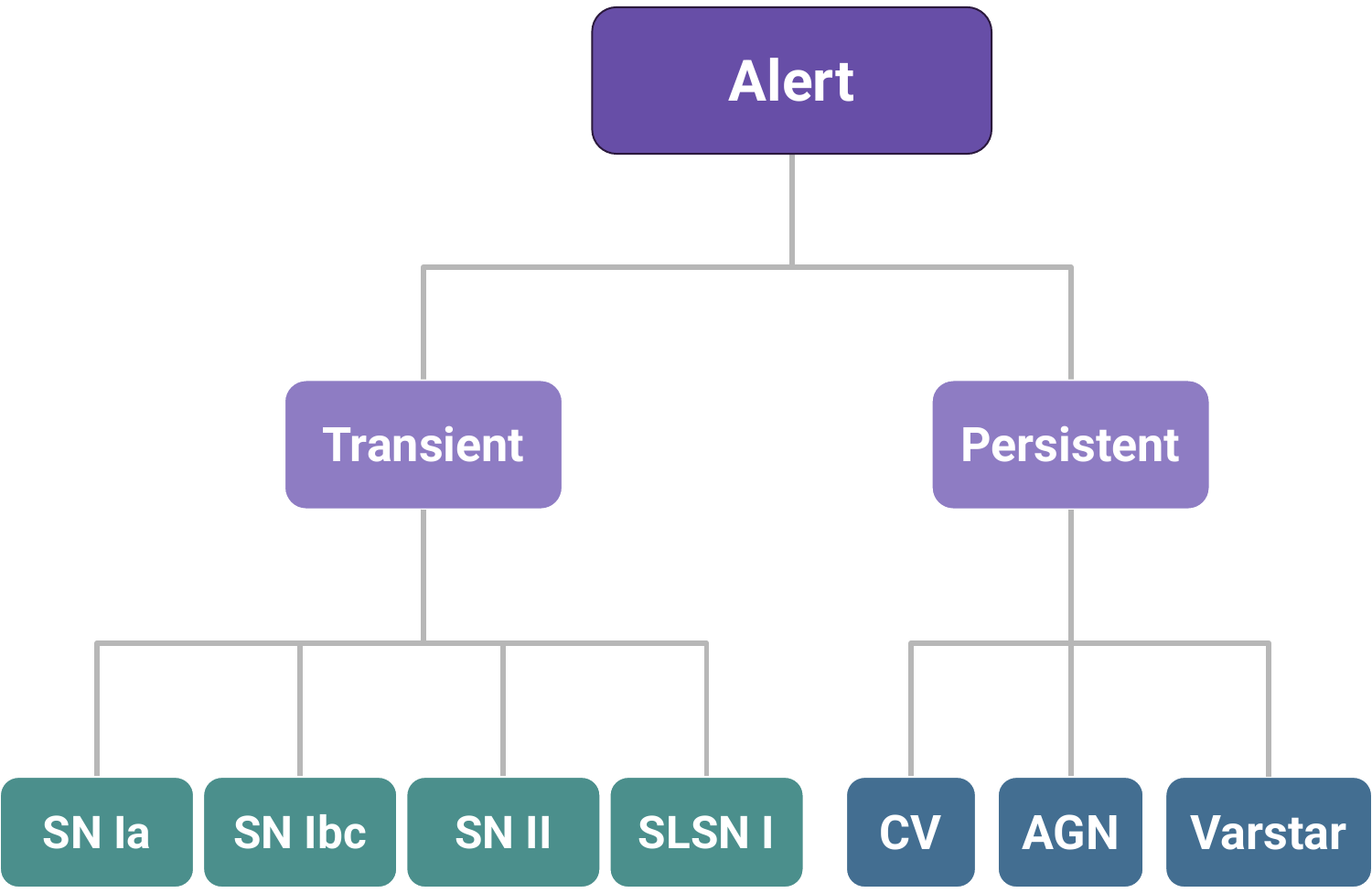}
    \caption{Taxonomy used for classifying sources from the Bright Transient Survey dataset.}
    \label{fig:bts_taxonomy}
\end{figure}

In addition to constructing our dataset, we organize our class labels along a 2-layer, observationally motivated taxonomy as shown in Figure \ref{fig:bts_taxonomy}. This enables us to perform hierarchical classification by explicitly defining the astrophysical relation between different classes. As discussed in \cite{oracle}, the exact choice of taxonomy depends on the science one hopes to achieve with the classifier. Our taxonomy design was motivated primarily by the available training data, our core science goals, and the discriminative power of the available modalities. For instance, we do not attempt to distinguish between stripped-envelope SN subtypes (e.g., SNe Ib versus SNe Ic), as such distinctions typically require spectroscopic observations rather than photometric or imaging data alone.

For our models, the first level performs binary classification between transient and persistent sources and is intended to be used as a discovery engine to triage and follow up new extragalactic transients, similar to \texttt{BTSbot} \citep[][]{BTSBot1, Rehemtulla+25_Pretraining}. Meanwhile, the level 2 performs a 7-way classification, which is comparable to a more general light curve classifier such as \texttt{AppleCiDEr} \citep{apple_cider_25} or \texttt{Superphot+} \citep{superphot}. As we will discuss in Section \ref{sec:conclusion}, the hierarchical classification framework we develop as part of this work is flexible and can accommodate different choices of taxonomies, allowing for adaptation to several different science cases.

\subsection{LSST (ELAsTiCC)}
\label{sec:dataset:ELAsTiCC}

The ``Extended LSST Astronomical Time-series Classification Challenge" \citep[\texttt{ELAsTiCC\footnote{\url{https://portal.nersc.gov/cfs/lsst/DESC\_TD\_PUBLIC/ELASTICC/}};}][]{ELAsTiCC, Malz23_ELAsTiCC, Malanchev23_ELAsTiCC, Knop23_ELAsTiCC} is a simulated dataset for LSST containing 32 distinct models of transient and variable astrophysical phenomena. It contains $\sim$50 million observations of $\sim$5 million individual sources in the LSST \textit{ugrizY} bands. While we do not discuss this dataset in great detail here, we refer interested readers to Section 3 of \cite{oracle} for more information about the dataset, the class mappings, and the taxonomy used to train the models. Table \ref{table:elasticc_featues} contains detailed information about the features used for the new (\base) \texttt{ELAsTiCC} models. 

\section{Models and Architectures}
\label{sec:arch}
\begin{figure}
    \centering
    \includegraphics[width=\linewidth]{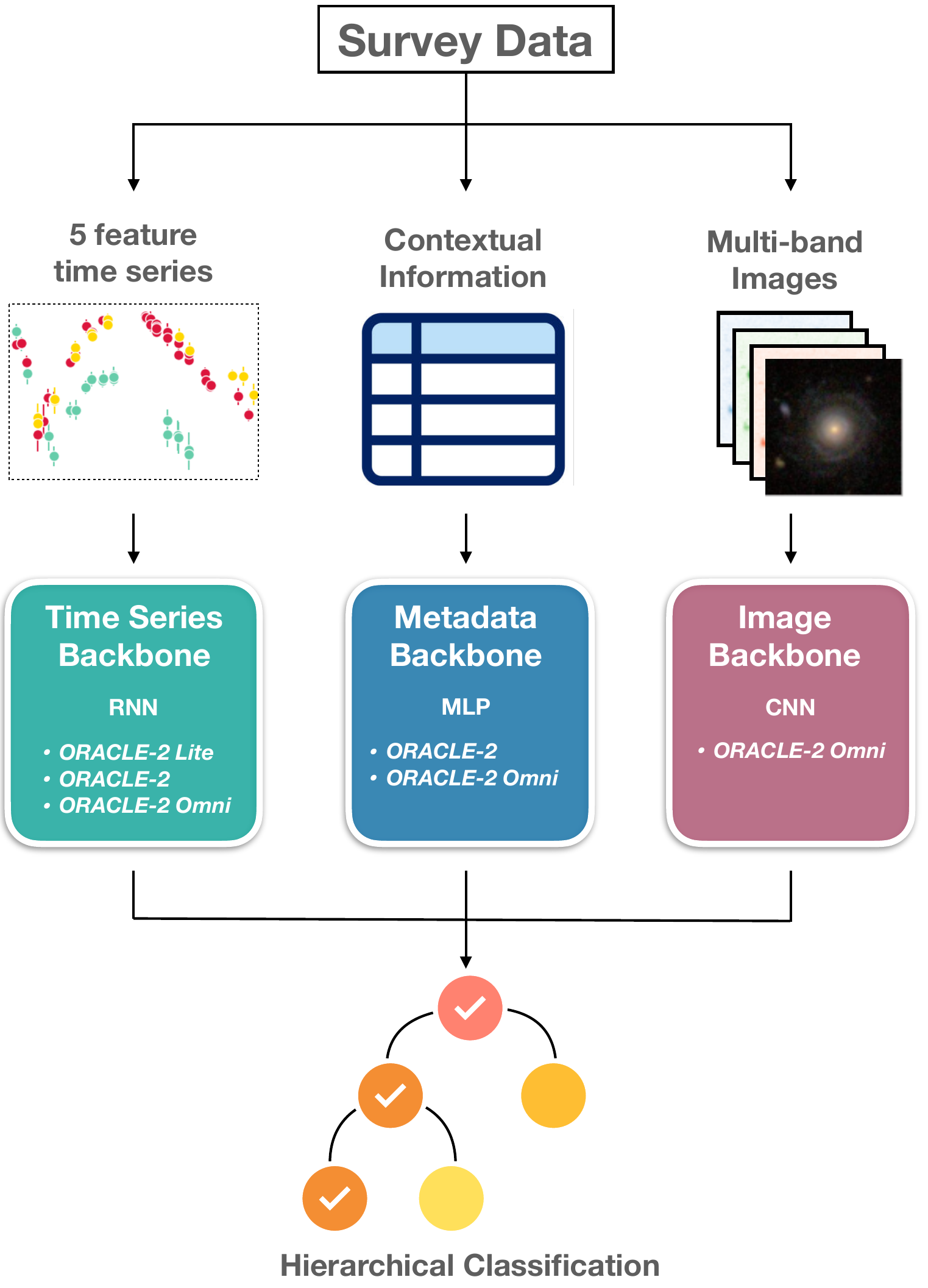}
    \caption{High level overview of the architecture and I/O for the \base\ family of models. The time series backbone uses a recurrent neural network (RNN), the metadata backbone uses a multi-layer perceptron (MLP), and the image backbone uses a convolutional neural network (CNN).}
    \label{fig:backbones}
\end{figure}

Incorporating multiple input modalities requires deliberate design choices in both the network architecture and the training routines. In this section, we introduce three new models: \lite, which performs classification using light curves only; \base, which combines light curves with tabular metadata; and \omni, which integrates light curves, metadata, and images. Figure \ref{fig:backbones} provides an overview of our model design. All three models share common architectural backbone(s), with additional components introduced only when necessary to accommodate new data modalities. This design decision enables a controlled assessment of how including new modalities affects the performance of our models. We note here that we only train the \omni\ model for ZTF since the \texttt{ELAsTiCC} dataset does not contain images. 

\subsection{Oracle-2 Lite}
\label{sec:arch:lite}

Dozens of machine-learning models have been developed for light curve classification \citep[][among others]{Rapid, avacado, supernnova, SuperRAENN20, Gomez2020_fleet, scone, parsnip21, Gagliano_2023_shallow, superphot, sheng24_needle, tdescore, ATAT, Cartagena25_vit_lc, apple_cider_25, ATCAT, oracle, starembed, Townsend26}. Broadly, these models adopt one of two design paradigms: 
\begin{enumerate}
    \item \textit{Feature-based methods} which extract hand-crafted features from the observations either by fitting the data to an analytical model (e.g., \citealt{superphot}) or by using statistical properties of the light curves (e.g., \citealt{avacado}). These features are then used as inputs for classical models such as random forests or gradient boosted decision trees.
    \item \textit{End-to-end deep-learning methods} which can natively handle sequential data and rely on the training process to learn salient features directly from the observations. Popular choices for this approach in astronomy involve the use of recurrent neural networks (RNN) \citep[e.g., ][]{Rapid, oracle} and transformers \citep[e.g., ][]{ATAT, ATCAT}. 
\end{enumerate}

While both approaches have shown promise and are used in production systems, deep learning allows models to learn abstract features directly from the data—features that can be difficult to design or engineer by hand. Such architectures also provide a natural framework for integrating multiple input modalities into a single system through the use of modality-specific backbones. Furthermore, \cite{ATAT} compared a random forest model with handcrafted features \citep[based on models from][]{Sanchez-Saez21_alerce} to modern deep-learning architectures on the simulated \texttt{ELAsTiCC} dataset and found that the deep-learning approach achieved meaningfully better performance (between $+5\%$ and $+10\%$ macro F1-score). 

For these reasons, we adopt a deep-learning approach where the models learn important representations from the data without explicit feature engineering. Specifically, we use Gated Recurrent Units \citep[GRUs, ][]{GRU}, a type of RNN, since they can natively handle arbitrary length sequences and have demonstrated strong performance in the classification of transient astrophysical phenomena \citep{Rapid, Chaini2020_RNN, Gagliano_2023_shallow, oracle}. We represent the time series input as follows.

Each observation $i$ in the time series is represented as a feature vector $X_i = [t_i, m_i, \epsilon_i, \lambda_i, f_i]$, where: 
\begin{enumerate}
    \item $t_i$ is the time, in days, since first detection
    \item $m_i$ is the brightness of the source, in AB mags for the BTS models and in flux for the \texttt{ELAsTiCC} models.
    \item $\epsilon_i$ is the $1\sigma$ uncertainty on the brightness
    \item $\lambda_i$ is the mean channel wavelength (in $\mu m$)
    \item $f_i$ is the detection flag, where $f_i$=1 for detections and $f_i$=0 for non-detections
\end{enumerate}
For a source $s$ with $N_s$ time steps, we represent the time series as a $N_s\times 5$ matrix. Since $N_s$ can differ for each source, we first pad our matrices using \texttt{pad\_sequence} and then pack them into batches using \texttt{pack\_padded\_sequence} within \texttt{PyTorch} \citep{pytorch}. This enables us to collate multiple light curves into a single batch during training and inference. 

Compared to \texttt{ORACLE-1}, the \base\ models introduce a new attention pooling mechanism over the entire light curve sequence, based on work by \cite{Bahdanau14_attention_pooling}. This mechanism aggregates a variable-length sequence of hidden states $H = \{h_1, h_2,...,h_T\}$ from the GRU into a fixed-dimensional context vector $c$. Specifically, given a GRU hidden state for time step $t$, $h_t \in R^{256}$, the mechanism computes the attention scores ($e_t$)

\begin{equation}
    e_t = v^T\text{tanh}(Wh_t + b),
\end{equation}

where matrix $W$ and vector $v$ are learned parameters. Then, we apply a softmax over the attention scores of all the hidden states to obtain the weight ($a_t$) for each hidden state

\begin{equation}
    a_t = \frac{\text{exp}(e_t)}{\sum_{i=1}^{T}\text{exp}(e_i)}.
\end{equation}

Finally, we use the weights to compute a weighted sum of our hidden states which gives us the context vector for a source

\begin{equation}
    c = \sum_{i=1}^{T} a_ih_i
\end{equation}

This mechanism allows the model to combine different parts of the input when identifying features relevant for classification, rather than relying solely on the final hidden state of the GRU, as in \texttt{ORACLE-1}. The goal is to mitigate a common limitation of RNNs which is their difficulty in capturing long-term dependencies as sequence length increases, ultimately degrading downstream classification performance (see Section \ref{sec:results}). These considerations are especially relevant for high-cadence, multi-year time-domain surveys such as LSST and the Argus Optical Array, where persistent sources can produce thousands of observations, resulting in very long light curves.

\begin{figure}
    \centering
    \includegraphics[width=\linewidth]{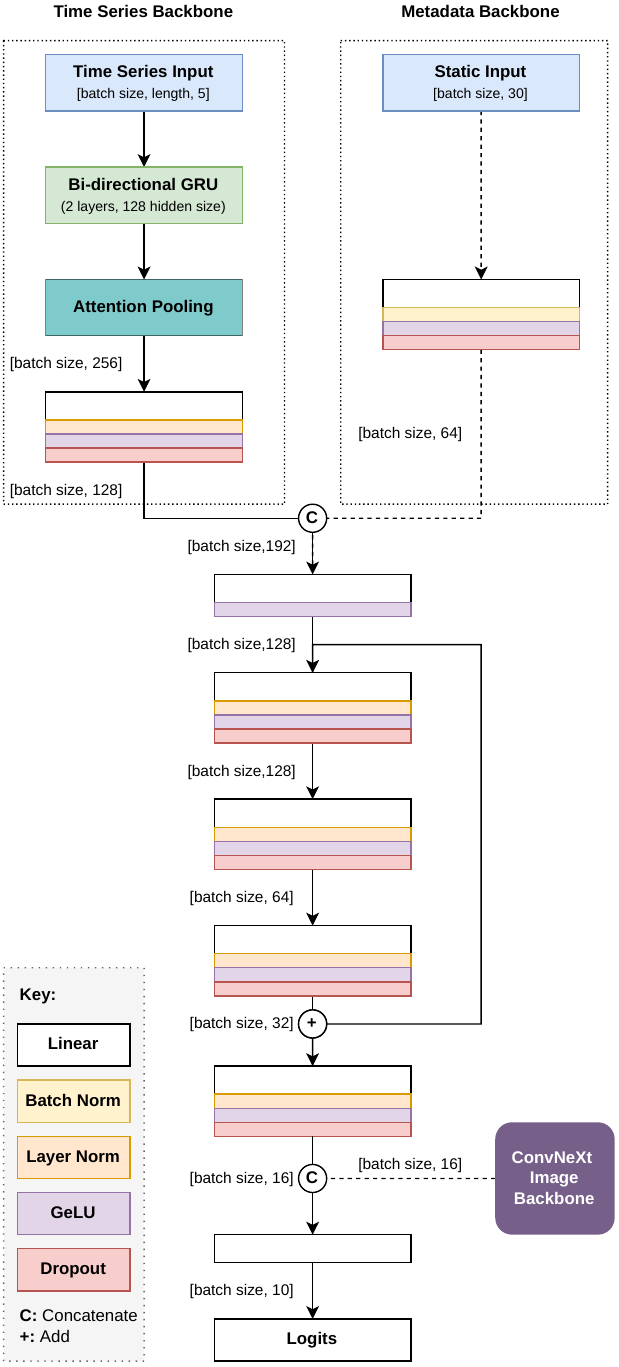}
    \caption{Detailed overview of the architecture for all three BTS models developed as part of this work. Connections common to all three models are shown with solid lines. Dashed lines show optional connections such as the metadata backbone (for the \base\ and \omni) and image backbone (for the \omni\ models).}
    \label{fig:architecture}
\end{figure}

These enhancements, combined with the use of a bidirectional GRU, better normalization, and improved training routines (Section \ref{sec:training}) \citep{Residuals_16, GeLU} make the \base\ models more performant than \texttt{ORACLE-1} across all classification metrics (Section \ref{sec:results}). A detailed schematic of the architecture, including details omitted here for brevity, is shown in Figure \ref{fig:architecture}. This network forms the time series backbone used for the entire \base\ family of models. Not only are these models effective for classification given enough training data (see Section \ref{sec:results}), but they also act as strong starting points to train models for other surveys via transfer learning \citep{Gupta25_transfer-learning}.

\subsection{Oracle-2}
\label{sec:arch:base}

In addition to light curves, alert packets from large time-domain surveys contain a lot of relevant information in the form of tabular metadata. While the specific contents of the metadata vary by survey (see Section \ref{sec:dataset} for more information), they typically contain information such as the on-sky location of the source, properties of other nearby sources, and host galaxy information (if available). Adding these features to the time-evolving information encoded within the light curve can help contextualize the astrophysical origin of the source and potentially improve the classification performance of a model. 

To incorporate this information, we use a simple multilayer perceptron \citep[MLP;][]{Rumelhart1986_backprop} and concatenate the embeddings with the light curve representation produced by the time series backbone. A complete list of the metadata used for the BTS and the \texttt{ELAsTiCC} models can be found in Table \ref{table:bts_featues} and Table \ref{table:elasticc_featues} respectively. Exact details about the metadata MLP architecture, such as the dimensionality of the layers, are specific to the dataset and do not materially affect the core model design. Figure \ref{fig:architecture} shows the architecture of this input branch for the BTS dataset. This forms the metadata backbone for both \base\ and \omni\ models.

\subsection{Oracle-2 Omni}
\label{sec:arch:omni}

In addition to the time-evolving emission of the source, details about the environment in which an event occurs are highly correlated with its astrophysical origin, as discussed in Section \ref{sec:dataset:BTS}. The presence of such correlations motivates our decision to include a third backbone to the \omni\ model to ingest images. 

Recent studies have shown that using off-the-shelf models pretrained on large amounts of data can deliver noticeable gains in classification performance over models trained from scratch. Somewhat surprisingly, this appears to be the case irrespective of whether the model was trained on astrophysical or terrestrial data \citep[such as ImageNet,][]{Imagenet}, albeit to varying degrees of success \citep{Cartagena25_vit_lc, Rehemtulla+25_Pretraining}. \cite{Walmsley+2024} tested 5 families of computer vision models, trained on $\sim$842,000 annotated Galaxy Zoo \citep{Lintott+2008, Lintott+2011} images from five observatories and found that the \texttt{ConvNeXt} \citep{ConvNext} models performed the best. These results were further corroborated for transient classification by \cite{Rehemtulla+25_Pretraining} who found that the \texttt{ConvNeXt} models perform comparably to state-of-the-art vision transformer models, while allowing for much higher throughput on identical hardware. Motivated by these findings, we use an off-the-shelf convolutional neural network \citep[CNN;][]{LeCun1989_cnn}, the \texttt{ConvNeXt-Pico}. Specifically, we use \texttt{Zoobot} variants which were trained to learn galaxy morphology by being trained to predict the volunteers' answers to questions posed by Galaxy Zoo (see \citealt{Walmsley+2024} for details). A detailed discussion of the architectural design involves engineering and machine learning considerations that are beyond the scope of this paper. We refer interested readers to \cite{ConvNext} for more information. 

Training this model correctly for our task requires a multi-step routine described in Section \ref{sec:training}; however, we make no major modification to the architecture itself besides adding a small MLP head for classification. This MLP head steps the dimensionality of the feature space from $512\rightarrow256\rightarrow64\rightarrow16$, after which the embeddings are concatenated with the other branches. See Figure \ref{fig:architecture} for more details.

\section{Training}
\label{sec:training}
In this section, we detail the data-augmentation strategies (Section \ref{sec:training:augmentation}), the hierarchical loss function (Section \ref{sec:training:loss}),  modality specific considerations (Section \ref{sec:training:modality}), and hyperparameter tuning (Section \ref{sec:training:hpo}) used to train the \base\ models.

\subsection{Data Augmentation} 
\label{sec:training:augmentation}

To train the model for real-time lightcurve classification, we apply a transformation to the dataset in which each light curve is truncated to only include observations within $t$ days of the first detection, where $t=2^n$ and $n\sim \mathcal{U}([0,10])$. This results in time horizons ranging from 1 to 1024 days with a strong bias toward earlier phases to help with early time performance, forcing the model to learn to make predictions from partial light curves at varying evolutionary stages. At the start of each training epoch, a new value of $n$ is independently sampled for every light curve. In effect, the model is exposed to a diverse distribution of temporal contexts, improving its robustness for real-time deployment.

For the validation and test sets, we adopt a deterministic evaluation scheme where each light curve is truncated at fixed time horizons of $t$ days since the first detection, where $t = 2^{n}, \forall n \in \{0, 1, 2, ..., 10\}$. This enables us to evaluate performance consistently across several phases of evolution and to monitor the time-averaged validation loss. During training, model checkpoints are selected based on performance aggregated across all validation epochs, ensuring strong performance throughout the light-curve evolution rather than at a single epoch.

For the CNN models, each image is randomly rotated by $d$ degrees, where $d\sim\mathcal{U}(\{0, 90, 180, 270\})$. This augmentation aims to introduce rotational invariance for the image backbone and helps prevent overfitting. We further augment our images via random vertical and horizontal flips with a probability of $0.5$ each. These augmentations improve generalization while preserving the underlying astrophysical content of the images.

\subsection{The Loss Function} 
\label{sec:training:loss}

Given the hierarchical nature of the classification problem we are trying to address with these models, we opt to use a bespoke loss function. Specifically, we use the weighted hierarchical cross entropy loss function \citep[WHXE; ][]{WHXE}, based on the hierarchical cross entropy \citep[HXE; ][]{HXE} loss. Previous studies have demonstrated that this objective function is effective for training models to achieve strong performance on both real \citep{Villar25_Hierarchical_Classification} and simulated datasets \citep{oracle}. We formulate the WHXE loss function as follows:

The probability of class $C$ in our modified WHXE loss hierarchy can be formulated as
\begin{equation}
    p(C) = \prod_{l=0}^{h(C)-1} p (C^{(l)} \text{ }|\text{ } C^{(l+1)}), 
\label{eq:conditional-prob}
\end{equation}
where $p (C^{(l)} \text{ }|\text{ } C^{(l+1)})$ is obtained by applying hierarchical softmaxes \citep{HSoftmax} to sets of siblings in the taxonomy, $h(C)$ is the height of node C in the hierarchy, and  $C^{(0)},...,C^{(H)}$ represents the path from the root (at $C^{(H)}$) to the leaf (at $C^{(0)}$). 

Next, an additional term $\lambda$ is used to weigh the losses at different nodes, based on where they appear in the hierarchy: 
\begin{equation}
    \lambda (C^{(l)}) = exp (-\alpha \cdot d(C)),
\end{equation}
where $d(C)$ represents the depth of the node $C$ in the hierarchy, and $\alpha$ is a free parameter adjusting the priority of different levels in the hierarchy, tuned during hyperparameter optimization (see Section \ref{sec:training:hpo}). 

Unifying the HXE and WHXE losses, we can define a weight term to address class imbalance as
\begin{equation}
    W (C^{(l)}) =  \left(\frac{N}{N_{\text{nodes}} \cdot N_{\text{c}}} \right)^{\gamma},
\end{equation}
where $N$ is the total number of samples in the dataset, $N_{\text{nodes}}$ is the number of unique classes, $N_{\text{c}}$ is the number of samples of class $C$, and $\gamma$ is a parameter dictating the strength of the class weighting: $\gamma=1$ reduces to inverse (linear) class weighting, as formulated in \cite{WHXE}; while $\gamma=0$ reduces to the unweighted HXE loss, as formulated in \cite{HXE}. Since we want our models to perform well on minority classes, we choose to use $\gamma=1$ throughout this work.

Putting everything together, our loss function can be formulated as follows:
\begin{equation}
    \mathcal{L}_{\text{WHXE}} (p, C) = - \sum_{l=0}^{h-1} W (C^{(l)}) \lambda (C^{(l)}) \text{ log } p (C^{(l)} | C^{(l+1)}),
\end{equation}
where C is the node of the true class. 

This formulation allows us to optimize for the value of a single, differentiable function during training while maintaining the aforementioned benefits of hierarchical classification.

\subsection{Modality Specific Considerations}
\label{sec:training:modality}

Training multimodal networks with several input backbones proves to be challenging since the different model architectures we employ (such as RNNs vs CNNs) prefer fundamentally different hyperparameters (see Table \ref{tab:hp_tuning}). To remedy this, we followed a three-step training routine for the \omni\ models. 

First, we trained the \base\ model and the image backbone independently using the best hyperparameters for each. Then we use these two pre-trained models to construct the \omni\ model and freeze their weights while training a small MLP head (see Figure \ref{fig:architecture}) that concatenates their outputs and performs the hierarchical classification. This effectively turns the frozen models into feature extractors.\footnote{We find that using same embedding dimension for both branches ensured that one of the branches did not dominate the final output of the model.} This step allows the MLP to effectively combine the embeddings from the two branches to minimize the loss while maintaining training stability. We refer to this step as the warm-up stage. Finally, we unfreeze the entire network, allowing it to train all parameters with $10\%$ of the original learning rate. We find that this routine was effective at maintaining training stability while delivering large performance improvements. While this resulted in strong performance gains over several different metrics (see Section \ref{sec:result:classification}), we note that this approach may not be unique and different training ``recipes" may achieve similarly strong performance. 

\subsection{Hyperparameter tuning}
\label{sec:training:hpo}

All models were trained using the \texttt{Adam} optimizer \citep{Kingma2014AdamAM} with a scheduler that reduces the learning rate (lr) when the loss begins to plateau (\texttt{ReduceLROnPlateau} scheduler within \texttt{PyTorch}). Specifically, the scheduler waits for $20$ epochs after the last decrease in loss before reducing the lr, has a decay factor of $0.8$, and reaches a minimum lr that is $1\%$ of the original value. Each model was trained for a maximum of 1000 epochs with an early stopping criterion applied on the phase averaged F1 score of the validation set, which has a patience of 100 epochs with a minimum improvement value of $10^{-3}$. For all of our models, we track the macro F1 score averaged across all epochs and use the checkpoint that achieved the highest value on this metric with the validation set.

The lr, batch size, and $\alpha$ values for the BTS were chosen using a Bayesian hyperparameter optimizer implemented within \texttt{Weights and Biases} \citep{wandb}. Table \ref{tab:hp_tuning} shows all the hyperparameters that were explored for training each of our new models, with the best values highlighted in bold.

For the \texttt{ELAsTiCC} models, we find that several different combinations of reasonable hyperparameter choices converge to nearly identical performance. Thus, to ensure a fair comparison with earlier models, we use the same hyperparameters that were used to train the \texttt{ORACLE-1} and \texttt{ORACLE-1 Lite} models. We trained 5 different instances of each model, initialized with different weights, to account for the variance from the stochastic training process\footnote{At times, the standard deviation is 0 up to 2 decimal places. We report these as is to keep formatting consistent throughout.}.

\section{Results}
\label{sec:results}
\begin{table}[]
    \centering
    \begin{tabular}{l|cc}
    \toprule
    \toprule
     & \multicolumn{2}{c}{\textbf{Image Backbone F1}} \\
     & \textbf{ZTF 1-Channel} & \textbf{PS1 3-Channel} \\
    \midrule
    \midrule
    \multicolumn{3}{c}{\textbf{Depth 1}} \\
    \midrule
    Persistent & 0.86±0.00 & \textbf{0.88±0.01} \\
    Transient & 0.92±0.00 & \textbf{0.93±0.00} \\
    \midrule
    \textbf{accuracy} & 0.90±0.00 & \textbf{0.91±0.00} \\
    \textbf{macro} & 0.89±0.00 & \textbf{0.91±0.00}\\
    % weighted avg & 0.90±0.00 & 0.91±0.00 \\
    \midrule
    \midrule
    \multicolumn{3}{c}{\textbf{Depth 2}} \\
    \midrule
    AGN & 0.82±0.01 & \textbf{0.85±0.01} \\
    CV & 0.70±0.01 & \textbf{0.80±0.02} \\
    Varstar & 0.80±0.02 & \textbf{0.87±0.01} \\
    SN II & 0.36±0.02 & \textbf{0.41±0.04} \\
    SN Ia & 0.67±0.03 & \textbf{0.75±0.02} \\
    SN Ib/c & \textbf{0.11±0.02} & \textbf{0.09±0.01} \\
    SLSN I & \textbf{0.09±0.06} & \textbf{0.09±0.12} \\
    \midrule
    \textbf{accuracy} & 0.64±0.02 & \textbf{ 0.71±0.01} \\
    \textbf{macro} & 0.51±0.01 & \textbf{0.55±0.02} \\
    % weighted avg & 0.65±0.01 & 0.71±0.01 \\
    \bottomrule
    \bottomrule
    \end{tabular}
    \caption{Depth 1 and depth 2 F1 scores for the single channel ZTF and triple channel Pan-STARRS-1 Image Backbones. The best performance (within $1\sigma$ uncertainties) on each metric is highlighted in bold.} 
    \label{tab:img_model_comparison}
\end{table}

We present performance on both the BTS (Section \ref{sec:result:classification:bts} and \ref{sec:result:classification:img}) and \texttt{ELAsTiCC} (Section \ref{sec:result:classification:ELAsTiCC}) datasets. In Section \ref{sec:results:throughput} we discuss the throughput for each of our models and consider the performance-throughput tradeoffs. Section \ref{sec:results:failures} discusses common failure modes for our models.

\subsection{Classification Performance}
\label{sec:result:classification}

First, we define the precision and recall for a class as Precision = TP/(TP+FP) and Recall = TP/ (TP+FN), where TP, FP, and FN are the number of true positives, false positives, and false negatives, respectively. Typically, we want to balance both metrics since high precision enables us to do targeted follow-up of rare or scientifically valuable sources, while high recall enables complete rate studies. The F1 score is the harmonic mean of the precision and recall and is formulated as $\text{F1 score} = 2 \cdot \text{Precision} \cdot \text{Recall}/\text{(Precision + Recall)}.$

For the rest of the discussion, we will focus on the macro-averaged F1 score, which is the mean of the F1 scores computed for each class in our dataset. We use the macro F1 score since it balances precision and recall, \textit{and} is sensitive to performance on minority classes. This is especially important for highly imbalanced datasets, such as our BTS sample, where class imbalance between majority and minority classes exceeds $50:1$ (see Section \ref{sec:dataset:BTS}).

\subsubsection{Image Backbone Comparison on BTS}
\label{sec:result:classification:img}

\begin{figure}
    \centering
    \includegraphics[width=\linewidth]{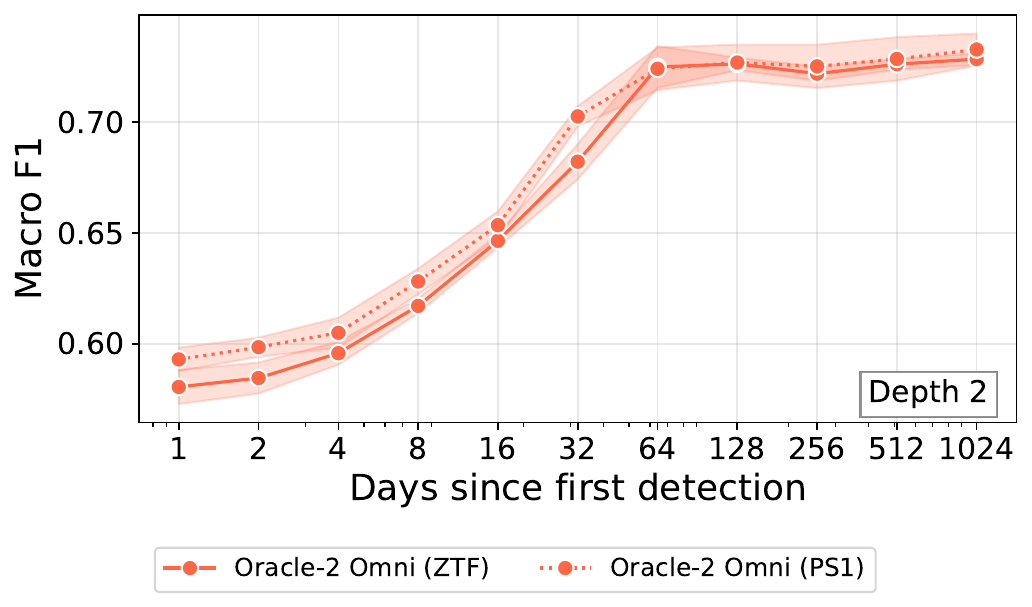}
    \caption{Depth 2 F1 scores as a function of time for the single channel ZTF and triple channel Pan-STARRS-1 \omni\ models. }
    \label{fig:omni_models_diff_branches}
\end{figure}

\begin{figure*}
    \centering
    \includegraphics[width=\linewidth]{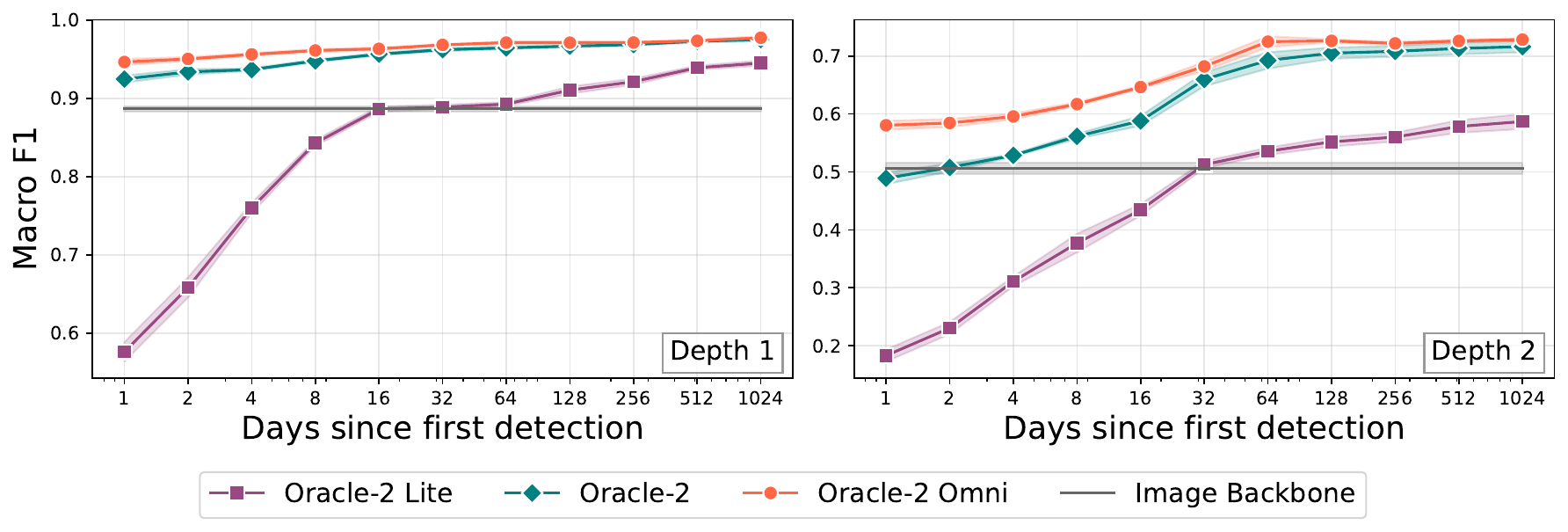}
    \caption{Time evolution of the macro F1 scores at depth 1 (left) and depth 2 (right) for the Image Backbone (image only), \lite\ (light curve only), \base\ (light curve + metadata), and \omni\ (light curve + metadata + images) models for the ZTF BTS dataset. Depth 1 distinguishes between transient and persistent sources while depth 2 allows for more granular classification between supernova sub-types, CVs, AGN, and variable stars.}
    \label{fig:bts_f1}
\end{figure*}

As discussed in Section \ref{sec:dataset}, we train our image backbone on both single-channel ZTF images as well as triple-channel PS images. See Figure \ref{fig:image_comparisons} for some examples showing differences between the two sets of images. We find that the image model trained on PS1 images performs at least as well as, and typically better than, the model trained on ZTF reference images. Table \ref{tab:img_model_comparison} shows the breakdown by class at both levels in the taxonomy. Based on these tests, we report a 4\% improvement in macro F1 score and a 7\% improvement in the accuracy with the model trained on the PS1 images when compared to the ZTF variant. 

In general, we find that image-only models demonstrate strong performance, not just for the classification of galactic vs extra-galactic sources, which may be easier due to the presence of crowded fields in the plane of the galaxy, but also for the different subtypes of SNe (namely core-collapse vs thermonuclear) and AGN. This implies that the models implicitly learn salient features of the host galaxy from image data alone, in contrast to existing approaches that rely on explicitly engineered host-galaxy features \citep{ghost, Villar25_Hierarchical_Classification}.

Interestingly, despite the gap in standalone image classification performance shown in Table \ref{tab:img_model_comparison}, we observe little difference between the corresponding \omni\ models at most epochs (see Figure \ref{fig:omni_models_diff_branches}). We hypothesize that the light curve morphology at late times and the color information available through the metadata compensates for the lack of multi-band imaging, allowing the \omni\ model equipped with the single-channel ZTF backbone to learn sufficiently rich representations for our classification task. Given the comparable performance of the \texttt{Omni} models, we proceed with the single-channel ZTF backbone for all subsequent experiments, as it is the more practical choice for real-time deployment with our broker.

\subsubsection{Performance on BTS}
\label{sec:result:classification:bts}

\begin{table*}[]
    \setlength{\tabcolsep}{3.5pt}   % default is 6pt
    \centering
    \begin{tabular}{l|ccc|ccc|ccc}
        \toprule
        \toprule
          & \multicolumn{3}{c|}{\lite} & \multicolumn{3}{c|}{\base} & \multicolumn{3}{c}{\omni} \\
         & $F1_{1}$ & $F1_{8}$ & $F1_{128}$ & $F1_{1}$ & $F1_{8}$ &  $F1_{128}$ &  $F1_{1}$ & $F1_{8}$ &  $F1_{128}$ \\
        \midrule
        \midrule
        \multicolumn{10}{c}{\textbf{Depth 1}} \\
        \midrule
        Persistent & 0.59±0.01 & 0.80±0.00 & 0.88±0.01 & 0.91±0.01 & 0.93±0.00 & \textbf{0.96±0.00} & \textbf{0.93±0.00} & \textbf{0.95±0.00} & \textbf{0.96±0.00} \\
        Transient & 0.56±0.02 & 0.88±0.00 & 0.94±0.00 & 0.94±0.00 & 0.96±0.00 & \textbf{0.98±0.00} & \textbf{0.96±0.00} & \textbf{0.97±0.00} & \textbf{0.98±0.00} \\
        \midrule
        \textbf{accuracy} & 0.58±0.01 & 0.85±0.00 & 0.92±0.00 & 0.93±0.00 & 0.95±0.00 & \textbf{0.97±0.00} & \textbf{ 0.95±0.00} & \textbf{0.96±0.00} & \textbf{0.97±0.00} \\
        \textbf{macro} & 0.58±0.01 & 0.84±0.00 & 0.91±0.00 & 0.92±0.00 & 0.95±0.00 & \textbf{0.97±0.00} & \textbf{ 0.95±0.00} & \textbf{0.96±0.00} & \textbf{0.97±0.00} \\
        % weighted avg & 0.57±0.01 & 0.85±0.00 & 0.92±0.00 & 0.93±0.00 & 0.95±0.00 & 0.97±0.00 & 0.95±0.00 & 0.96±0.00 & 0.97±0.00 \\
        \midrule
        \midrule
        \multicolumn{10}{c}{\textbf{Depth 2}} \\
        \midrule
        AGN & 0.56±0.01 & 0.72±0.01 & 0.75±0.02 & \textbf{0.91±0.01} & \textbf{0.95±0.01} & \textbf{0.96±0.00} & \textbf{0.92±0.01} & \textbf{0.94±0.01} & \textbf{0.96±0.00} \\
        CV & 0.30±0.01 & 0.52±0.03 & 0.72±0.01 & 0.75±0.01 & 0.81±0.01 & \textbf{0.91±0.01} & \textbf{0.82±0.01} & \textbf{0.86±0.00} &\textbf{0.90±0.00} \\
        Varstar & 0.16±0.01 & 0.26±0.02 & 0.41±0.01 & 0.84±0.03 & \textbf{0.88±0.01} & \textbf{0.90±0.02} & \textbf{0.89±0.01} & \textbf{0.90±0.01} & \textbf{0.91±0.01} \\
        SN II & 0.11±0.02 & 0.30±0.02 & 0.66±0.01 & 0.30±0.03 & 0.39±0.02 & \textbf{0.70±0.01} & \textbf{0.40±0.01} & \textbf{0.46±0.01} & \textbf{0.72±0.01} \\
        SN Ia & 0.10±0.02 & 0.65±0.03 & 0.87±0.01 & 0.44±0.05 & 0.64±0.02 & 0.88±0.01 & \textbf{0.66±0.02} & \textbf{0.76±0.01} & \textbf{0.90±0.00} \\
        SN Ib/c & 0.03±0.02 & 0.13±0.02 & 0.33±0.02 & 0.10±0.01 & 0.15±0.01 & \textbf{0.32±0.03} & \textbf{0.17±0.01} & \textbf{0.18±0.01} & \textbf{0.32±0.02} \\
        SLSN I & 0.00±0.00 & 0.04±0.03 & 0.13±0.03 & 0.07±0.01 & 0.10±0.01 & 0.28±0.05 & \textbf{0.20±0.02} & \textbf{0.23±0.02} & \textbf{0.37±0.05} \\
        \midrule
        \textbf{accuracy} & 0.26±0.01 & 0.53±0.02 & 0.73±0.01 & 0.54±0.03 & 0.65±0.02 & 0.84±0.01 & \textbf{0.67±0.01} & \textbf{0.74±0.00} & \textbf{0.87±0.00} \\
        \textbf{macro} & 0.18±0.01 & 0.38±0.02 & 0.55±0.01 & 0.49±0.01 & 0.56±0.00 & 0.71±0.01 & \textbf{0.58±0.01} & \textbf{0.62±0.00} & \textbf{0.73±0.00} \\
        % weighted avg & 0.23±0.02 & 0.57±0.02 & 0.76±0.01 & 0.56±0.03 & 0.69±0.01 & 0.85±0.01 & 0.69±0.01 & 0.75±0.00 & 0.87±0.00 \\
        \bottomrule
        \bottomrule
    \end{tabular}
    \caption{Per-class F1, macro F1, and accuracies for the \lite\ (light curve only), \base\ (light curve + metadata), and \omni\ (light curve + metadata + images) models, across both levels of the BTS taxonomy, at various phases of light curve evolution. The best performance (within $1\sigma$ uncertainties) on each metric is highlighted in bold.}
    \label{tab:f1_bts}
\end{table*}

\begin{figure*}
    \centering
    \includegraphics[width=\linewidth]{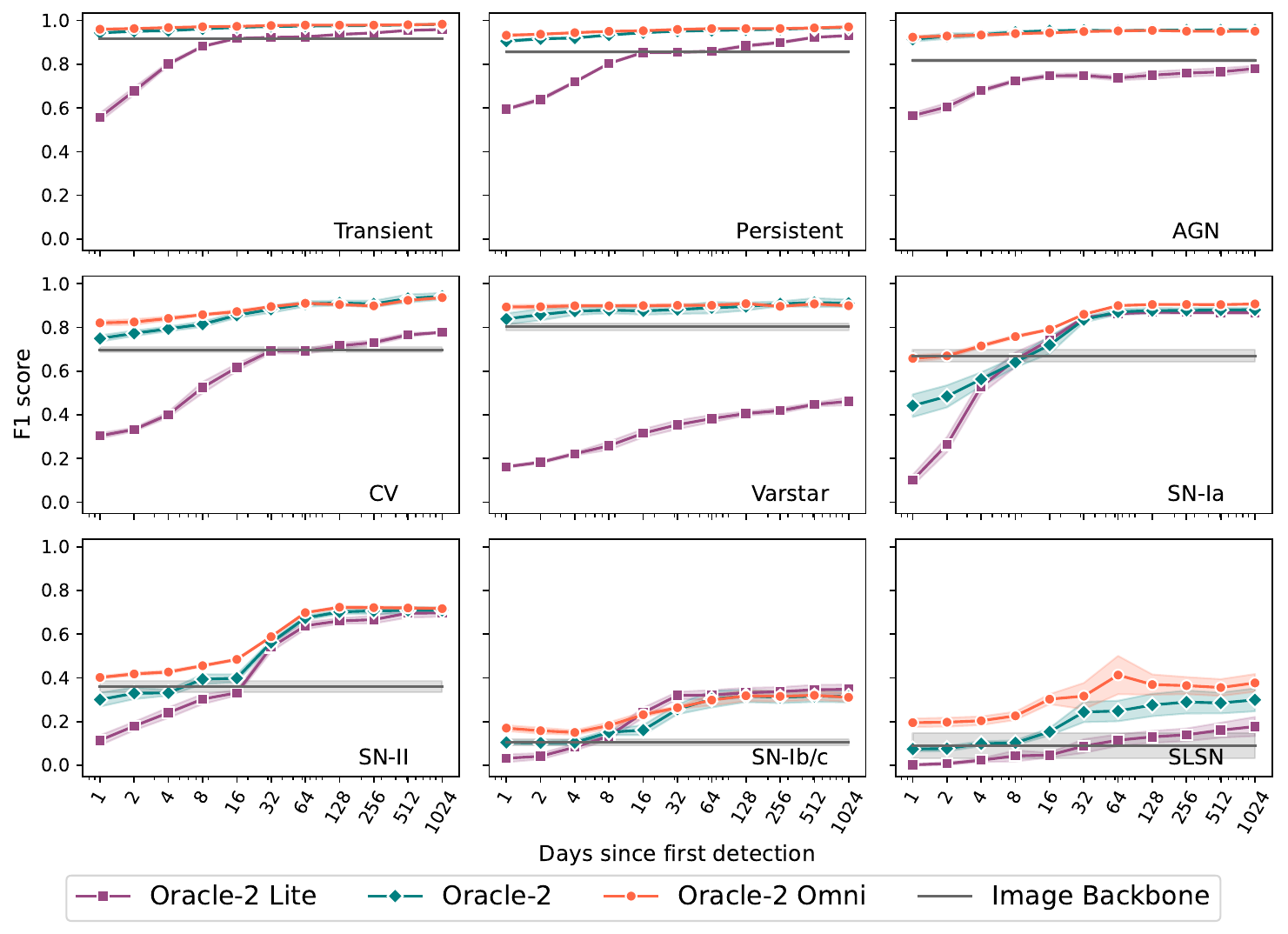}
    \caption{Time evolution of the class F1 scores across all 9 classes in our taxonomy for the Image Backbone (image only), \lite\ (light curve only), \base\ (light curve + metadata), and \omni\ (light curve + metadata + images) models for the Bright Transient Survey (BTS) dataset.}
    \label{fig:bts_f1_per_class}
\end{figure*}

\begin{figure}
    \centering
    \includegraphics[width=0.95\linewidth]{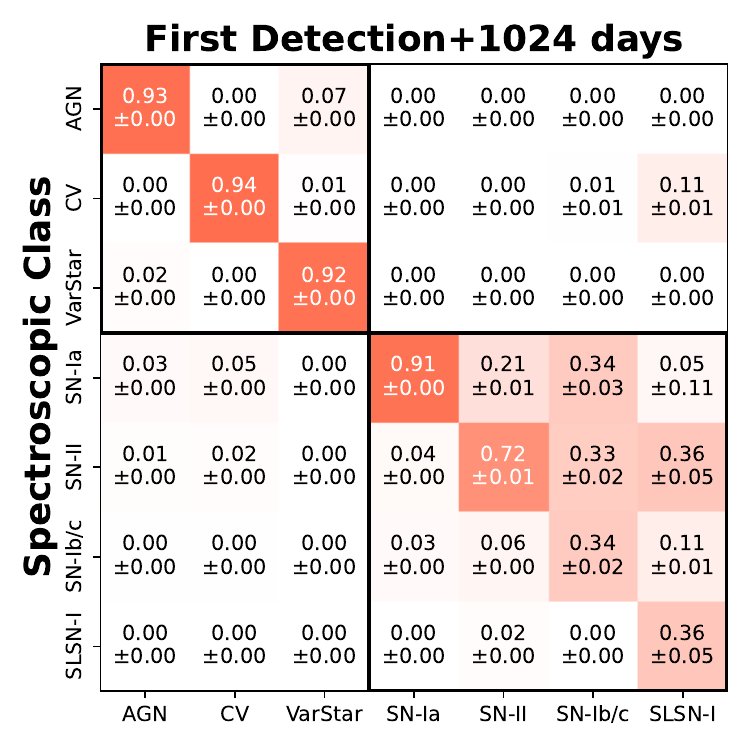}
    \includegraphics[width=0.95\linewidth]{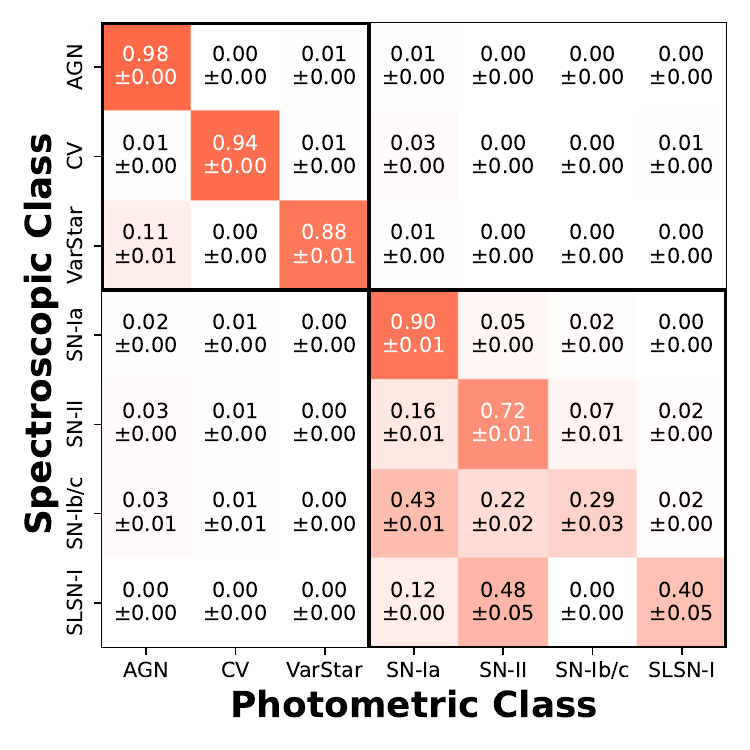}
    \caption{Depth 2 confusion matrices for the \omni\ model, 1024 days after the first detection, normalized by the predicted/photometric class (top) and the true/spectroscopic class (bottom).}
    \label{fig:bts_omni_f1}
\end{figure}

Figure \ref{fig:bts_f1} shows the time evolution of macro F1 scores on the BTS dataset for all three models at both levels in our taxonomy. Our results reveals a consistent ordering across all phases, at both levels of the taxonomy, with \omni\ achieving the highest scores, followed by \base, and then \lite. This difference in performance is especially pronounced at early phases, with the advantage diminishing, in both relative and absolute terms, at late times. Since adding the metadata branch has a negligible effect on throughput (see Section \ref{sec:results:throughput}) and is feasible from an infrastructure standpoint, we argue that it is always advantageous to use the \base\ model instead of \lite\ for the BTS models. Thus, we choose to focus the rest of this discussion on the differences between the \base\ and \omni\ models.

When considering the per-class performance, we find that adding modalities improves performance for every class in our taxonomy. Figure \ref{fig:bts_f1_per_class} and Table \ref{tab:f1_bts} show the time evolution of the F1 score for all three BTS models across all 9 classes in our taxonomy.  From Table \ref{tab:f1_bts}, we see significant gains in the class F1 scores for CVs (between $+0\%$ and $+7\%$), variable stars (between $+0\%$ and $+5\%$), SN\,Ia (between $+2\%$ and $+22\%$), SN\,Ib/c (between $+0\%$ and $+7\%$), SN\,II (between $+2\%$ and $+10\%$), and SLSN\,I\footnote{The large uncertainties in the SLSNe\,I metrics reflect the small number of SLSNe\,I examples in the test set.} (between $+0\%$ and $+13\%$) for the \omni\ model when compared to \base. We see smaller gains in performance for the Transient, Persistent, and AGN classes, possibly owing the informativeness of the other modalities or the relative simplicity of the classification task itself. 

For the macro F1 score, we see the largest difference in performance at early times with the \omni\ outperforming the \base\ model by between 9\% and 13\% in the first week. At later times, these advantages diminish as the light curve itself becomes increasingly discriminative. This suggests that multimodality is most valuable at early times, precisely when follow-up decisions are most difficult. We report a final depth 2 macro F1 score, at $1024$  days after the first detection, of $0.73\pm0.01$ and $0.71\pm0.01$ for the \omni\ and \base\ models, respectively. Figure \ref{fig:bts_omni_f1} shows the depth 2 confusion matrices for \omni\ at 1024 days after the first detection. For the accuracy metric, our \omni\ model outperforms \base\ by between $3\%$ and $13\%$, depending on the phase at which the models are evaluated. 

\begin{figure*}
    \centering
    \includegraphics[width=0.325\linewidth]{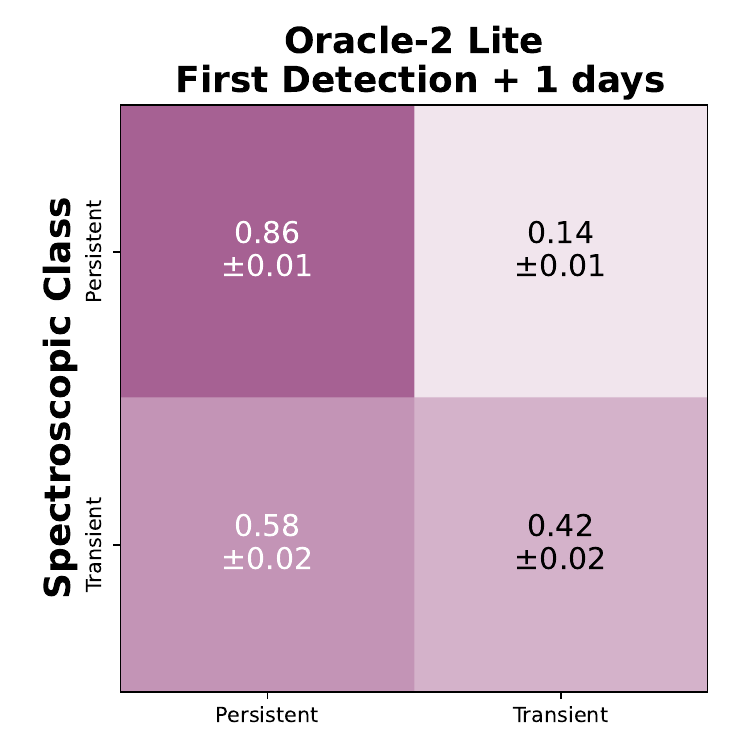}
    \includegraphics[width=0.325\linewidth]{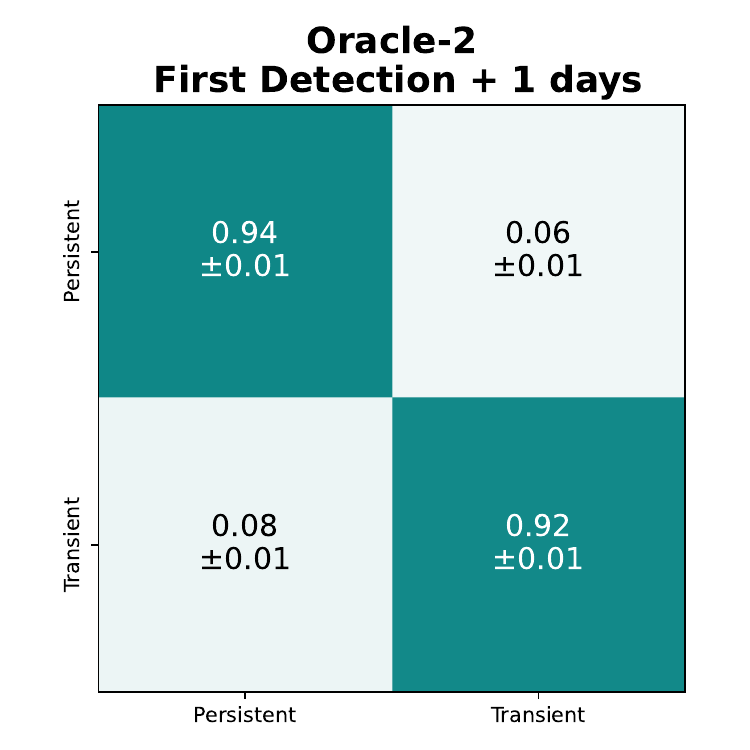}
    \includegraphics[width=0.325\linewidth]{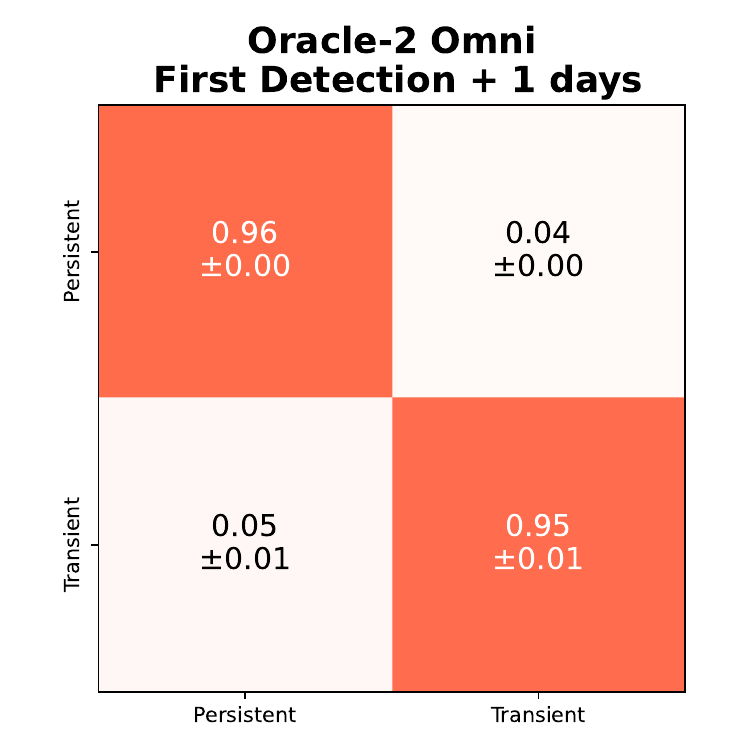}
    \includegraphics[width=0.325\linewidth]{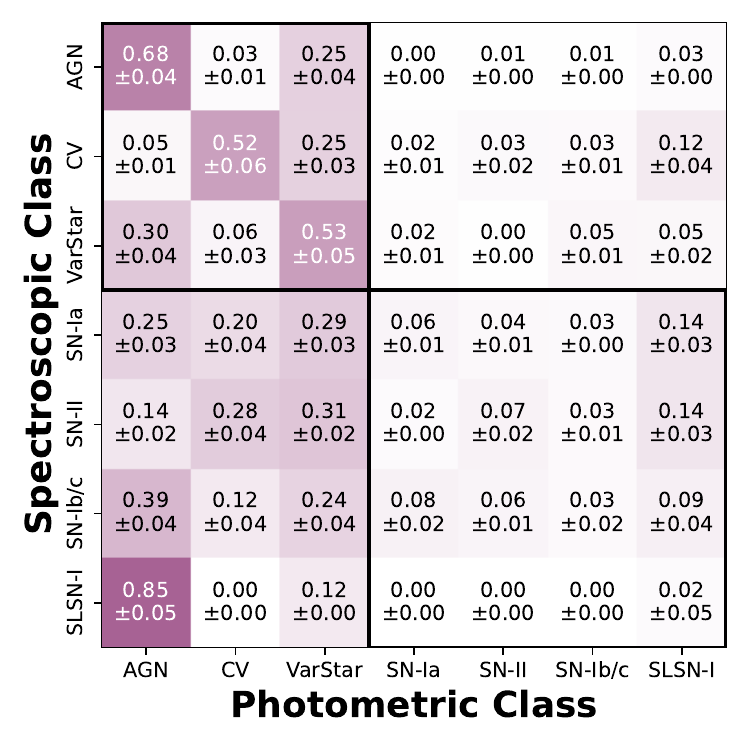}
    \includegraphics[width=0.325\linewidth]{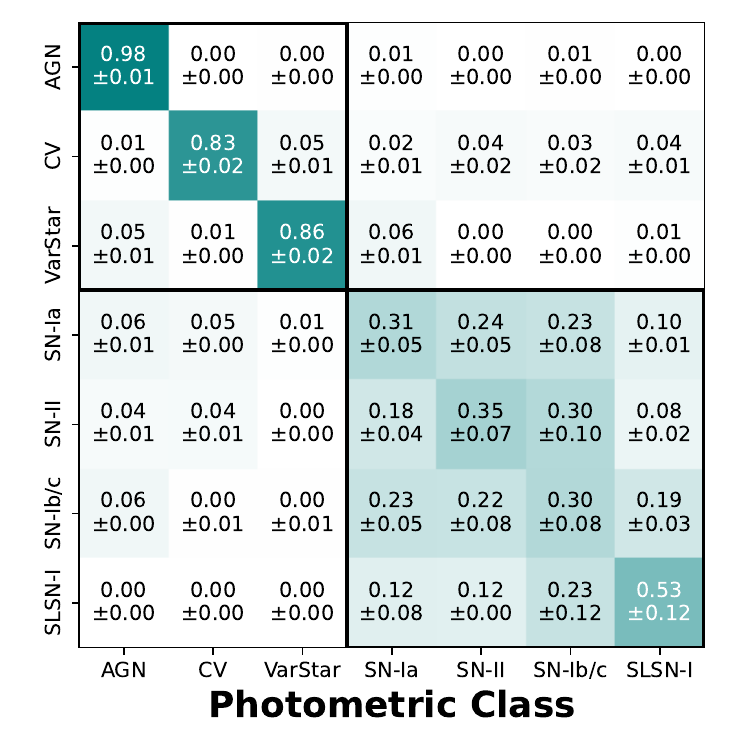}
    \includegraphics[width=0.325\linewidth]{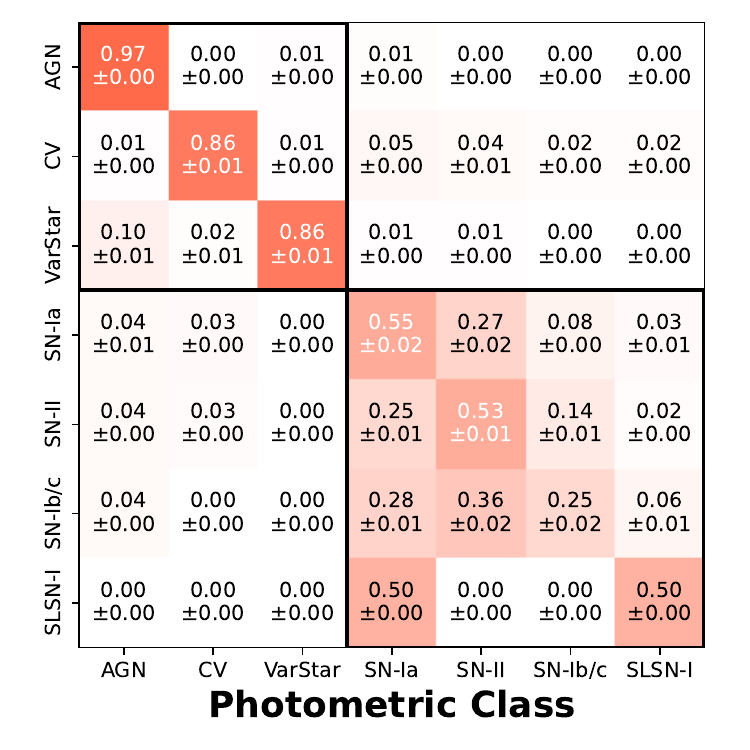}
    \caption{Depth 1 (top) and depth 2 (bottom) confusion matrices for the \lite\ (left; uses light curves only), \base\ (middle; uses light curves + metadata), and \omni\ (right; uses light curves + metadata + images) models, normalized by the true/spectroscopic class, 1 day after the first detection.}
    \label{fig:bts_cf_1_day}
\end{figure*}

\begin{figure*}
    \centering
    \includegraphics[width=0.32\linewidth]{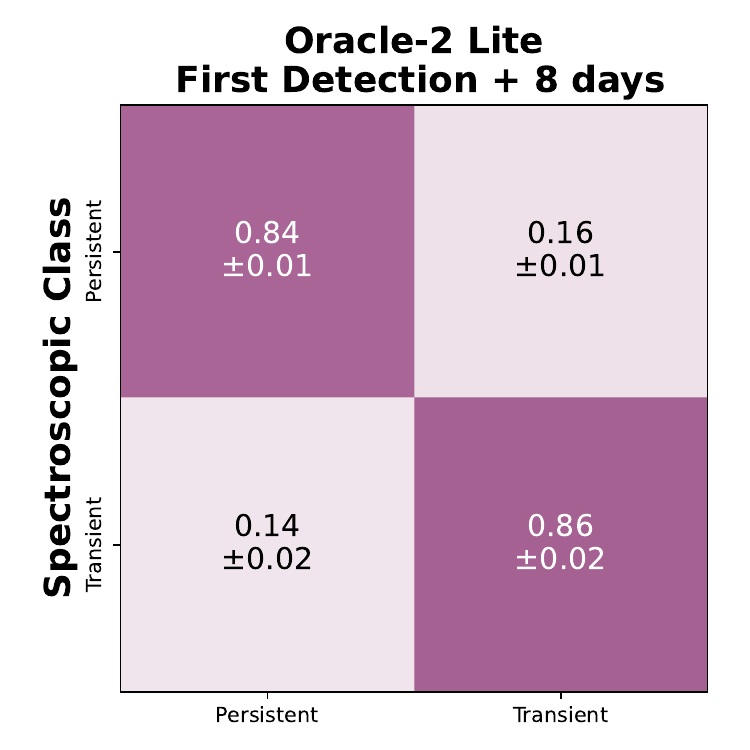}
    \includegraphics[width=0.32\linewidth]{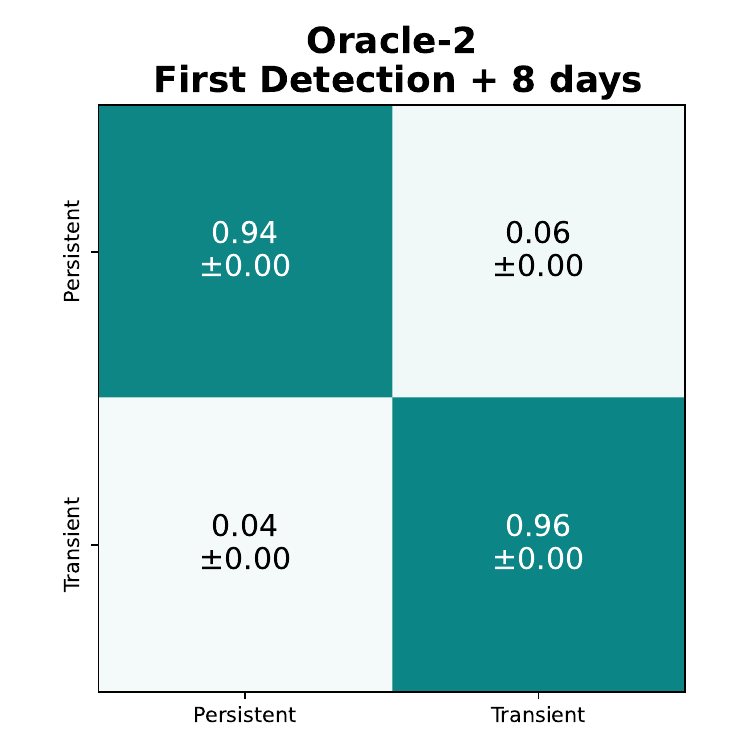}
    \includegraphics[width=0.32\linewidth]{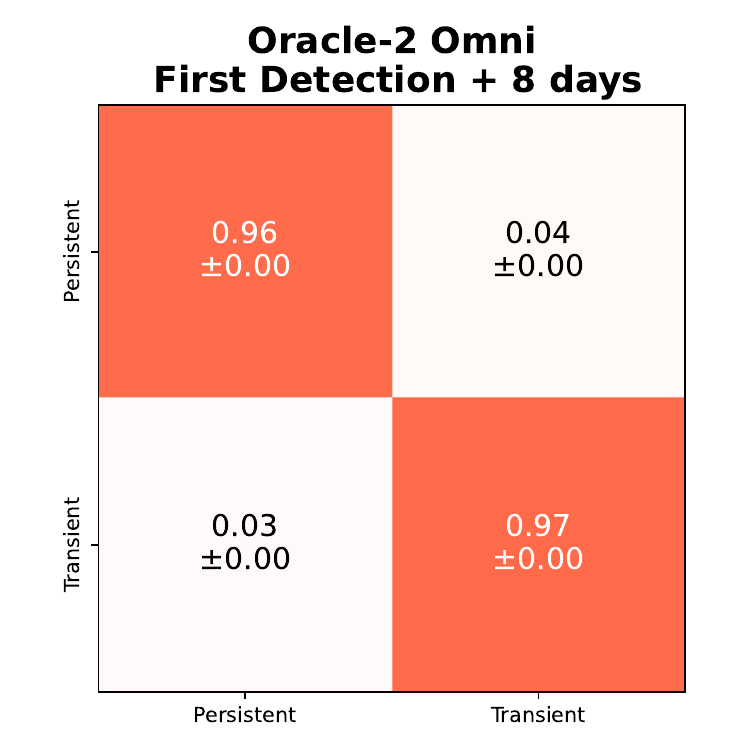}
    \includegraphics[width=0.32\linewidth]{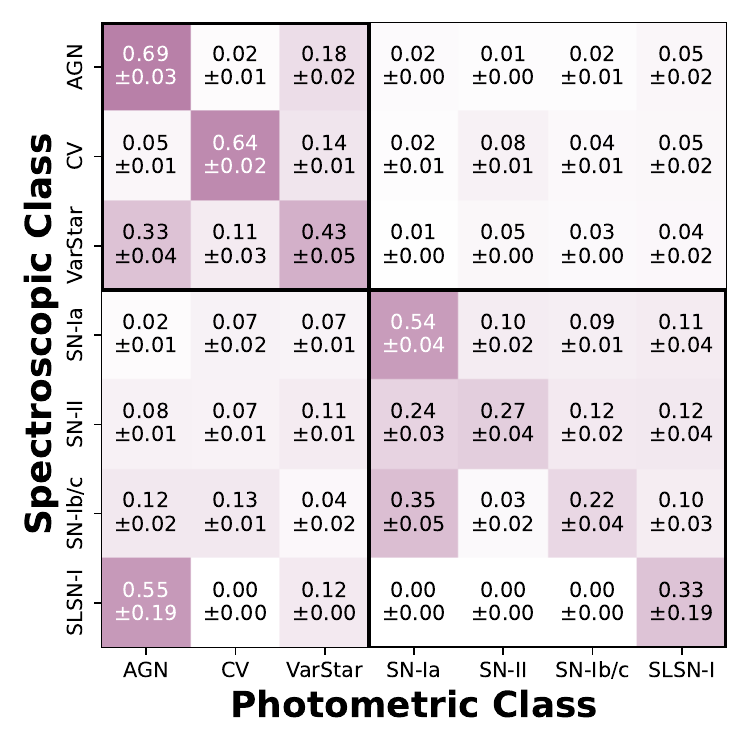}
    \includegraphics[width=0.32\linewidth]{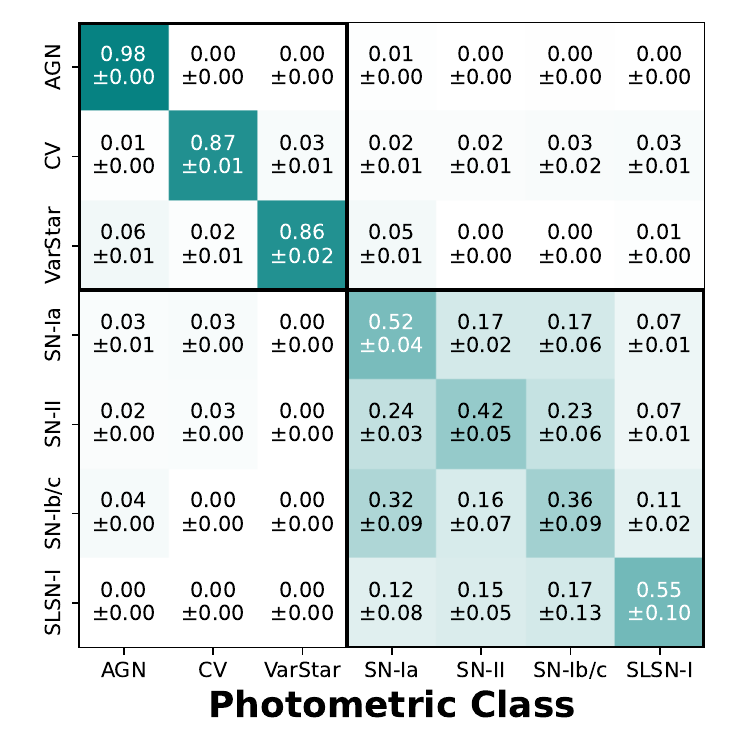}
    \includegraphics[width=0.32\linewidth]{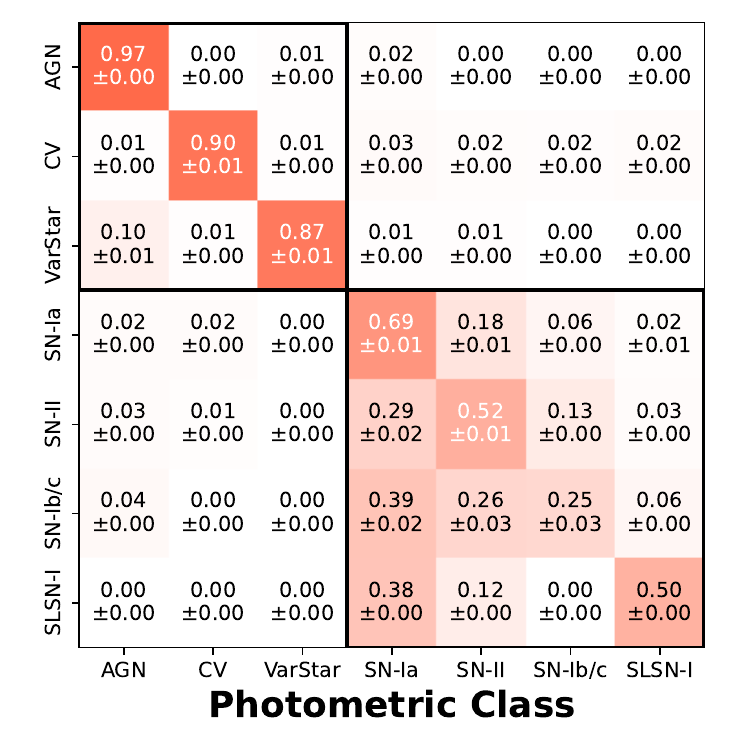}
    \caption{Depth 1 (top) and depth 2 (bottom) confusion matrices for the \lite\ (left; uses light curves only), \base\ (middle; uses light curves + metadata), and \omni\ (right; uses light curves + metadata + images) models, normalized by the true/spectroscopic class, 8 days after the first detection.}
    \label{fig:bts_cf_8_day}
\end{figure*}

\subsubsection{Performance on ELAsTiCC}
\label{sec:result:classification:ELAsTiCC}

\begin{figure*}
    \centering
    \includegraphics[width=\linewidth]{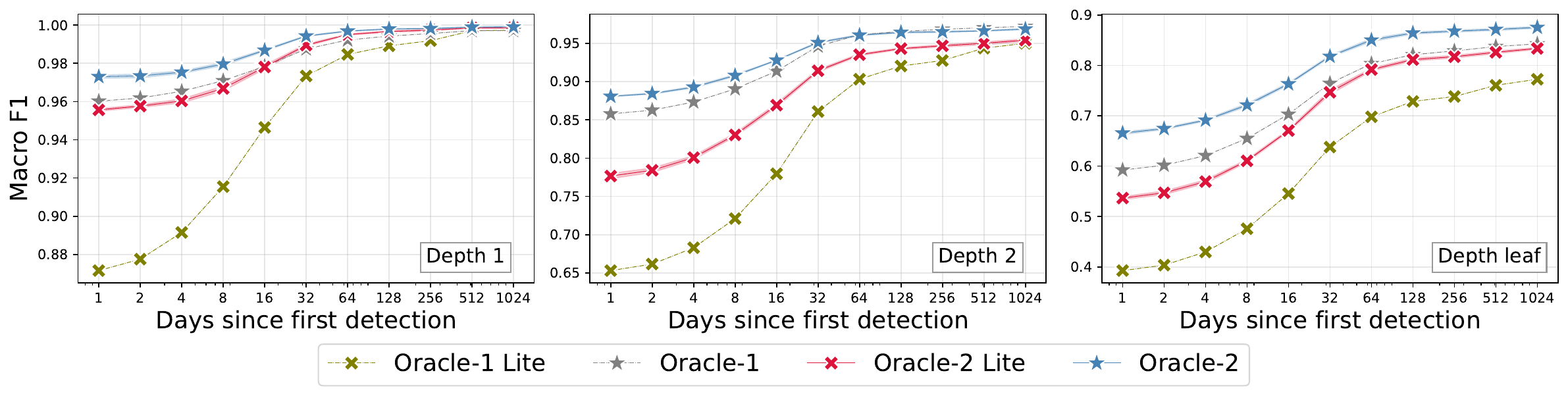}
    \caption{Time evolution of the macro F1 scores at depth 1 (left), depth 2 (middle), and leaf depth (right) for the \lite\ (light curve only) and \base\ (light curve + metadata) models for the \texttt{ELAsTiCC} dataset.}
    \label{fig:elasticc_f1}
\end{figure*}

\begin{figure*}
    \centering
    \includegraphics[angle=0, width=0.62\linewidth]{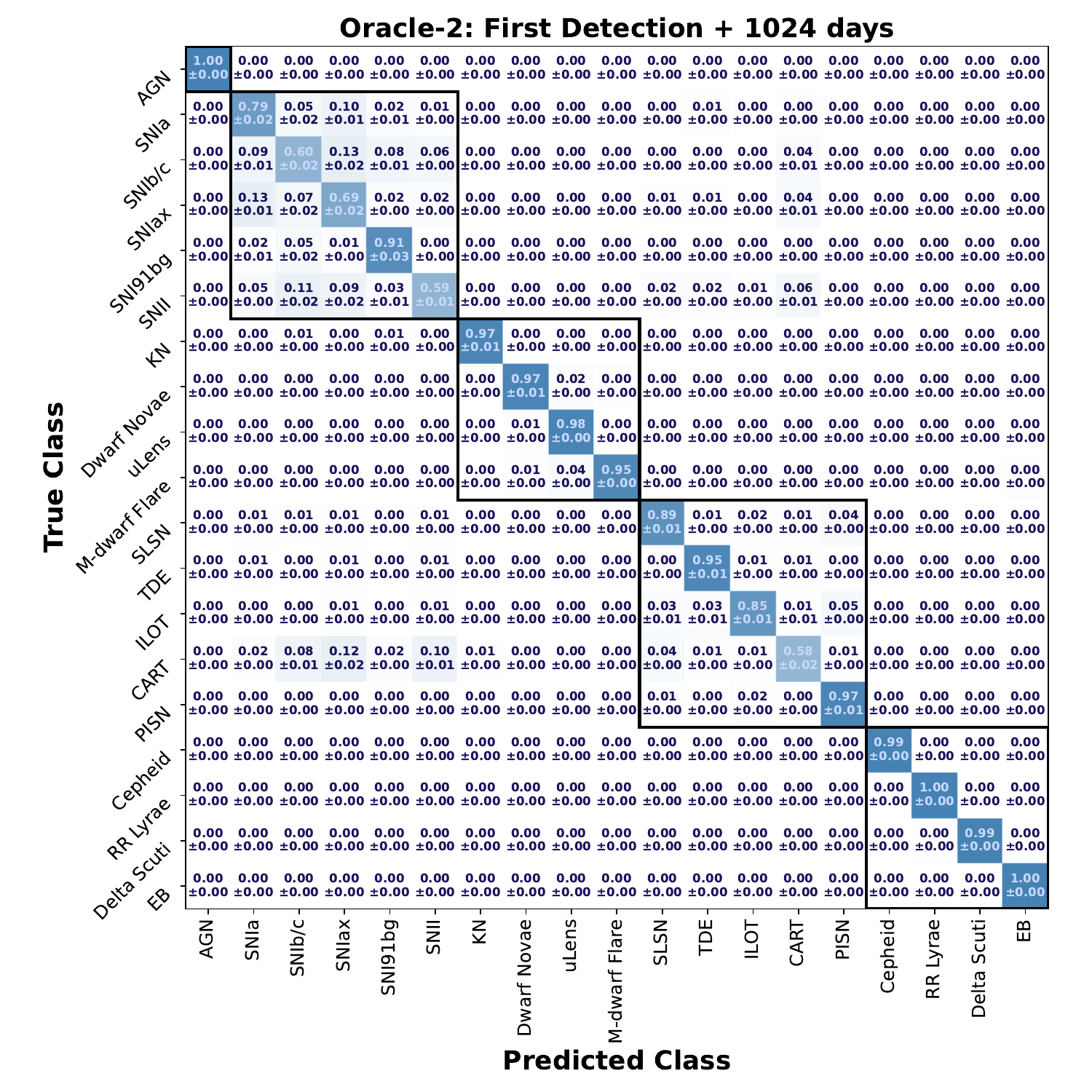}
    \includegraphics[angle=0, width=0.62\linewidth]{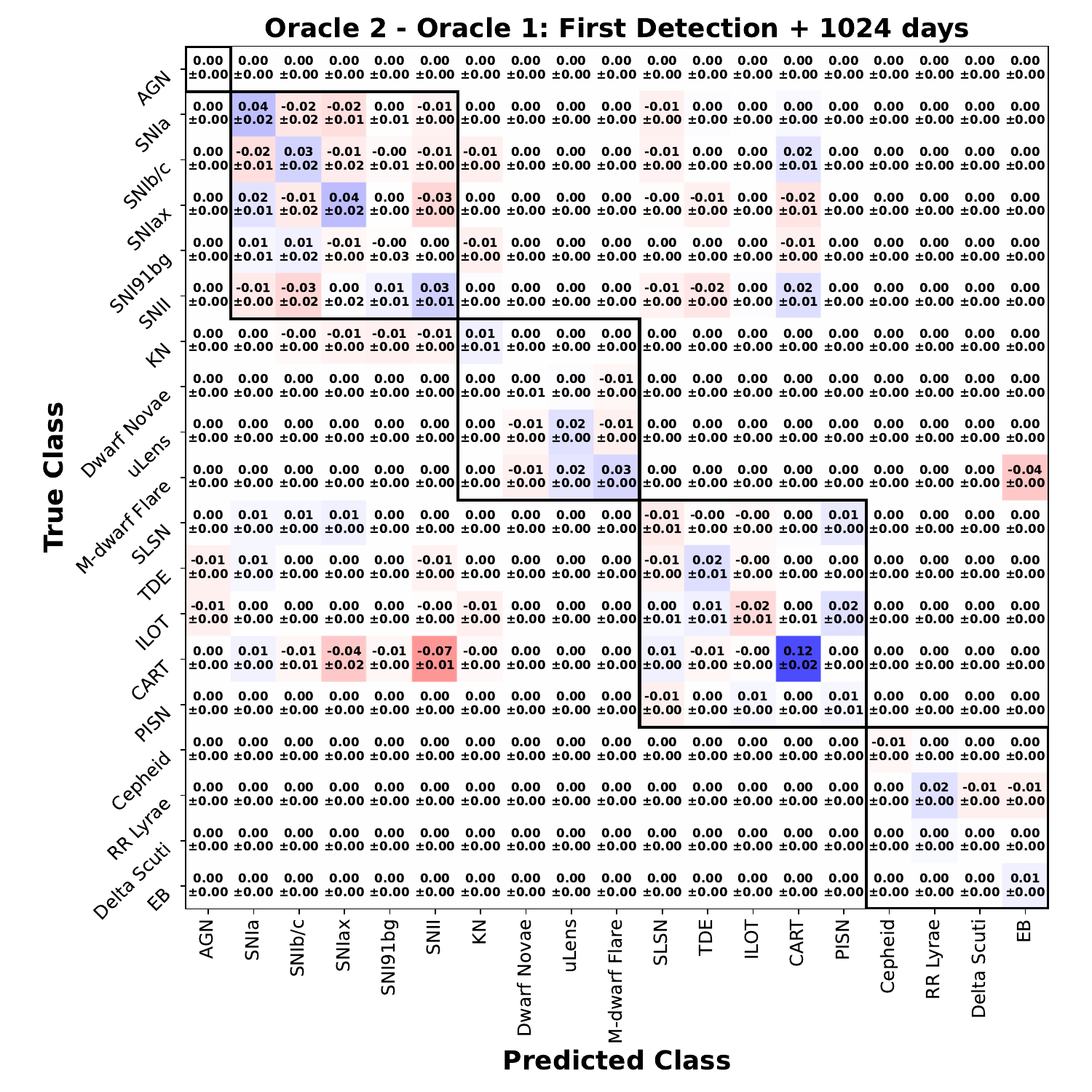}
    \caption{\textbf{Top:} Confusion matrix for the \base\ model trained on the \texttt{ELAsTiCC} dataset, 1024 days after the first detection. \textbf{Bottom:} Difference in the confusion matrices of the \base\ and \texttt{ORACLE-1} models, 1024 days after the first detection. \base\ consistently shows improvements in power (blue) along the diagonal elements and reduction in power (red) for off-diagonal elements, highlighting better agreement between the true and predicted labels for the \base\ models.}
    \label{fig:oracle2_elasticc_cf_1024}
\end{figure*}

We begin evaluating the \texttt{ELAsTiCC} models 1 day after the first detection. At this early phase, we report a top-level (depth 1) macro F1 score of $0.97\pm0.00$ for the \base\ model. As more observations are taken, it becomes possible to reliably perform classification at depth 2 with a macro F1 score $>0.90$, just 8 days after the first detection. At 64 days after the first detection, we report a macro F1 score $>0.80$ on the 19-way (leaf) classification task. While we will not discuss the performance at every phase in great detail here, we refer interested readers to Table \ref{tab:f1_elasticc} and Figure \ref{fig:elasticc_f1}, which shows the macro F1 score for our \texttt{ELAsTiCC} models as a function of time, for every depth in our taxonomy.

The \base\ and \lite\ models for \texttt{ELAsTiCC} reliably outperform their original counterparts at every phase across all levels of our taxonomy, as shown in Figure \ref{fig:elasticc_f1}. In fact, the new \lite\ model is able to perform comparably to the original \texttt{ORACLE-1} on the 19-way classification task at late times, despite not requiring any metadata. Given the strong performance of all models at depths 1 and 2 in the taxonomy, we choose to focus the rest of this discussion on the leaf depth performance of the models, where the differences are more pronounced. 

When we consider the per-class performance at the leaves, we report substantial improvements to the F1 scores for Calcium Rich Transients (CARTs), Intermediate Luminosity Optical Transients (ILOTs), and Kilonovae (KNe). While the \texttt{ORACLE-1} models do perform better on some classes, the margin is typically much smaller, resulting in a favorable macro-averaged outcome for the \base\ family. Figure \ref{fig:oracle2_elasticc_cf_per_class_f1} shows the evolution of the F1 score for each leaf class, as a function of time since first detection. These advantages are also apparent on the ``difference" confusion matrix shown in Figure \ref{fig:oracle2_elasticc_cf_1024}, which highlights how \base\ has better agreement along the diagonal and less power in the off-diagonal elements when compared to the original model, indicating much stronger agreement between the true and predicted classes. 

At 1024 days after the first detection, we report a 19-way (leaf depth) macro F1 score of $0.88\pm0.00$ and $0.83\pm0.00$ for the \base\ and \lite\ models, respectively. This brings our \base\ models in line with other state-of-the-art \texttt{ELAsTiCC} models such as \texttt{ATCAT} \citep[macro F1 $\sim0.89$,][most comparable to \base]{ATCAT} and \texttt{RoMAE} \citep[macro F1 $\sim0.80$,][most comparable to \lite]{Zivanovic25_Rope} and surpasing the performance of \texttt{ATAT} \citep[macro F1 $\sim0.84$,][]{ATAT} and the original \texttt{ORACLE-1} models (macro F1 $\sim0.84$). The final confusion matrix can be seen in Figure \ref{fig:oracle2_elasticc_cf_1024}.

\subsection{Throughput Performance}
\label{sec:results:throughput}

As the parameter count and complexity of our models continue to scale, practical considerations such as the throughput become increasingly important for real-time deployment. This problem is further exacerbated by the sheer volume of alerts that surveys such as LSST will produce. In this section, we report the throughput performance for each of our models to inform deployment decisions for current and future surveys.

Each model was tested with a batch size of one, with inputs consisting of light curves, postage stamps ($63\times63$ pixels, 3 channels), and dataset-specific metadata (30 features for BTS, 18 for ELAsTiCC). We use a batch size of one to faithfully represent our current deployment plan, where inference will be run every time an alert passes our filter (see Section \ref{sec:results:deployment}). Since the RNN inference time is a function of the sequence length, we use samples with 174 timesteps for the \texttt{ELAsTiCC} models and 41 timesteps for the BTS models, representing the median number of observations for a source in each dataset. For each model, the wall-clock inference time for 100 iterations was recorded and used to compute the mean and standard deviation of the throughput. All testing was completed on a system with an \texttt{Intel(R) Xeon(R) Gold 6230R} central processing unit (CPU) running at 2.10 GHz with 188 GB of memory, and an \texttt{NVIDIA A100} graphics processing unit (GPU) with 40 GB of video memory\footnote{We note that the throughput will depend on the hardware used.}. The inference throughput for each model using both the CPU and GPU, along with its parameter count, is reported in Table \ref{tab:throughput}.

\begin{table}[]
    \setlength{\tabcolsep}{4pt}
    \centering
    \begin{tabular}{l|cc|cc}
    \toprule
    \toprule
    \multirow{2}{*}{\textbf{Model}} &
    \multirow{2}{*}{\textbf{Dataset}} &
    \multirow{2}{*}{\textbf{N$_{\text{param}}$}} &
    \multicolumn{2}{c}{\textbf{Throughput (s$^{-1}$)}} \\
    &&& \textbf{CPU} & \textbf{GPU} \\
    \midrule
    
    \multirow{2}{*}{\lite} &
    \multirow{2}{*}{BTS} &
    \multirow{2}{*}{502K} &
    12.10 & 404.07 \\
    & & & $\pm$ 1.35 & $\pm$ 4.57 \\
    \midrule
    
    \multirow{2}{*}{\base} &
    \multirow{2}{*}{BTS} &
    \multirow{2}{*}{529K} &
    10.42 & 374.61 \\
    & & & $\pm$ 1.07 & $\pm$ 14.50 \\
    \midrule
    
    \multirow{2}{*}{\omni} &
    \multirow{2}{*}{BTS} &
    \multirow{2}{*}{9.2M} &
    7.67 & 115.48 \\
    & & & $\pm$ 1.10 & $\pm$ 0.43 \\
    \midrule
    
    \multirow{2}{*}{Image Branch} &
    \multirow{2}{*}{BTS} &
    \multirow{2}{*}{8.7M} &
    44.71 & 232.02 \\
    & & & $\pm$ 13.89 & $\pm$ 1.47 \\
    \midrule
    
    \multirow{2}{*}{\lite} &
    \multirow{2}{*}{\texttt{ELAsTiCC}} &
    \multirow{2}{*}{502K} &
    7.44 & 157.16 \\
    & & & $\pm$ 0.79 & $\pm$ 1.44 \\
    \midrule
    
    \multirow{2}{*}{\base} &
    \multirow{2}{*}{\texttt{ELAsTiCC}} &
    \multirow{2}{*}{529K} &
    6.53 & 150.87 \\
    & & & $\pm$ 0.86 & $\pm$ 3.37 \\
    
    \bottomrule
    \bottomrule
    \end{tabular}
    \caption{Inference throughput (s$^{-1}$) measured on a CPU (\texttt{Intel(R) Xeon(R) Gold 6230R}) and GPU (\texttt{NVIDIA A100}), together with the number of trainable parameters ($N_{\rm param}$), for each model in the \base\ family.}
    \label{tab:throughput}
\end{table}

Given the performance improvements we report for the \base\ models relative to \lite\ in Section \ref{sec:result:classification:bts}, we find that incorporating the metadata branch is always worthwhile, for both datasets. We make this recommendation since the large improvement in classification performance is coupled with a small difference in throughput. 

The trade-offs associated with the \omni\ model are similarly favorable. While incorporating the image branch reduces GPU throughput by approximately a factor of three on our test bench, the CPU throughput, which is more representative of our deployment environment, is only $\sim35\%$ percent lower than that of the \base\ model. Since our current production pipeline performs inference entirely on CPUs, the additional throughput penalty of the image branch is minimal in practice. We therefore conclude that the computational overhead of the \omni\ model is well justified by the improvement in classification performance. For applications in which inference is performed primarily on GPUs, however, the larger reduction in throughput may warrant a different trade-off between computational efficiency and classification performance.

Ultimately, the value derived from deploying \omni, and similar multimodal architectures, depends on the scientific objectives of the model, the operational constraints of a given broker, the underlying hardware used, and implementation details such as batch inference, queuing, and caching strategies. Thus, it is difficult to make universal recommendations.

\subsection{Failure mode analysis for BTS}
\label{sec:results:failures}

\begin{figure}
    \centering
    \includegraphics[width=\linewidth]{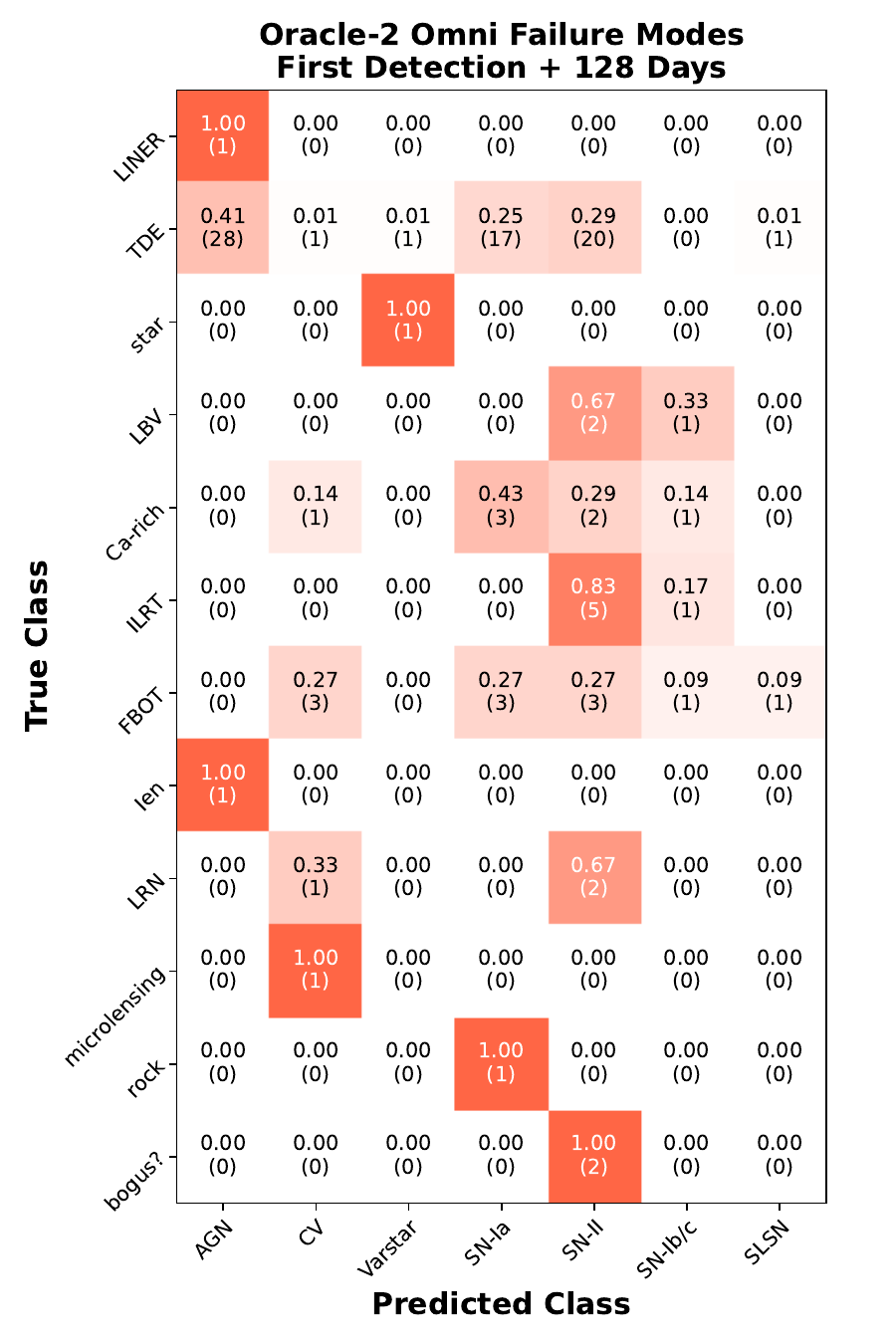}
    \caption{Asymmetric classification matrix, normalized by the true class, illustrating the classifications for sources which do not neatly fit into our taxonomy for BTS.}
    \label{fig:failure_mode}
\end{figure}

As we have stated in Section \ref{sec:dataset:BTS}, we do not apply any quality cuts to our data beyond what has already been filtered by BTS in order to maintain precision and recall metrics that are representative of real-world deployment scenarios. This results in a small fraction ($<1\%$) of sources that belong to classes which have very little training data and thus are not represented in our classification taxonomy. Some of these sources are peculiar subtypes of classes that are a part of our taxonomy, such as Ca-Rich Transient (which are likely SNe), while others belong to astrophysically distinct classes such as Tidal Disruption Events (TDEs). These sources can make our models fail in ways that are not represented in the ``traditional" confusion matrices shown in Section \ref{sec:result:classification:bts}.

Understanding how our models behave when presented with this data is crucial to ensure that we can make informed follow-up decisions and build robust systems around the model's output. Figure \ref{fig:failure_mode} shows an asymmetric classification matrix, illustrating how these sources are classified by the \omni\ model, 128 days after the first detection. While most of these classes have too few sources to make any strong statistical claims about the misclassifications, we can see that most rare transients are classified as different SN subtypes. By comparison, TDEs are most often misclassified as AGN, likely owing to their long light curves and nuclear origin, although they are also often classified as SN\,Ia or SN\,II. While there is no silver bullet to solving these failure modes, being aware of them allows us to better leverage the model and informs direction for future improvements.

\section{Real-time deployment}
\label{sec:results:deployment}
\begin{table}[]
    \centering

    \begin{tabular}{l|ccc|c}
    \toprule
    \toprule
     & \textbf{Precision} & \textbf{Recall} & \textbf{F1 Score} & \textbf{Count} \\
    \midrule
    \midrule
    \multicolumn{5}{c}{\textbf{Depth 1}} \\
    \midrule
    Persistent & 0.62 & 1.00 & 0.77 & 5 \\
    Transient & 1.00 & 0.99 & 1.00 & 339 \\
    \midrule
    \textbf{accuracy} &  &  & 0.99 &  \\
    \textbf{macro}  & 0.81 & 1.00 & 0.88 & 344 \\
    \textbf{weighted}  & 0.99 & 0.99 & 0.99 & 344 \\
    \midrule
    \midrule
    \multicolumn{5}{c}{\textbf{Depth 2}} \\
    \midrule
    AGN & 0.40 & 1.00 & 0.57 & 2 \\
    CV & 0.50 & 1.00 & 0.67 & 3 \\
    SLSN & 0.25 & 0.33 & 0.29 & 3 \\
    SN-II & 0.74 & 0.78 & 0.76 & 108 \\
    SN-Ia & 0.89 & 0.86 & 0.88 & 210 \\
    SN-Ib/c & 0.17 & 0.11 & 0.13 & 18 \\
    \midrule
    \textbf{accuracy} &  &  & 0.79 &  \\
    \textbf{macro} & 0.49 & 0.68 & 0.55 & 344 \\
    \textbf{weighted} & 0.79 & 0.79 & 0.79 & 344 \\
    \bottomrule
    \bottomrule
    \end{tabular}
    \caption{Performance metrics (precision, recall, F1 score, and accuracy) for the \omni\ model deployed on the live ZTF alert stream in the period between May 15 and June 11, 2026.}
    \label{tab:deployment_performance}
\end{table}

Based on the results from Section \ref{sec:results}, we have deployed the single-channel ZTF variant of the \omni\ model on the \texttt{BOOM} broker \citep{Jegou_du_Laz_25_Boom_Broker} after weighing the classification performance, follow-up latency requirements, and available compute.  In our production system, all inference is performed on CPUs, as they provide sufficient throughput for the current ZTF alert stream. The model runs on the filtered alert stream \citep[similar to the one detailed in][]{BTS2} provided by \texttt{BOOM}, consuming alerts via Kafka and uploading classifications to \texttt{Fritz} in near real-time. Inference is re-run on each new alert for a given source, keeping the class scores on \texttt{Fritz} continuously updated. After an initial period of testing and debugging, we deployed a stable version of the model on the ZTF alert stream on May 15, 2026. Between May 15 and June 11, 2026, we monitored the model's performance and compared its predictions against sources with spectroscopic classifications.

During this period, \omni\ classified 344 new sources\footnote{To ensure data hygiene, we only consider sources with a first alert occurring after 1 January, 2026 i.e. after the cutoff date for our original dataset.} which also had a spectroscopic label, achieving a depth 1 time-averaged macro F1-score of 0.88 and accuracy of 0.99. At depth 2, we report a time-averaged macro F1-score of 0.55 and accuracy of 0.79. Table~\ref{tab:deployment_performance} summarizes the statistics from the real-time deployment of \omni. While we lack sufficient data to produce results analogous to Figures~\ref{fig:bts_f1} and \ref{fig:bts_f1_per_class}, we report strong performance on SNe\,Ia (F1=0.88) and SNe\,II (F1=0.76). 

In addition to classifying sources which also had spectroscopic labels, \omni\ has produced photometric classifications for 1099 other sources. As is, the model has already been used for determining sources for spectroscopic follow-up and its results have been used for identifying rare and interesting transient sources \citep{2026TNSAN.201....1S}, weeks after deployment.

We expect some performance degradation relative to the test set since we are targeting fainter sources ($m>18.5$ mag) and due to recent changes in ZTF survey cadence \citep{ZTF_1day_cadence}, both of which introduce distribution shift; however, as shown here, this degradation has not been catastrophic. We intend to maintain stable performance through a combination of continuous monitoring and periodic retraining of the model.

\section{Discussion}
\label{sec:discussion}
Our results (Section \ref{sec:results}) highlight both the promise and the challenges of real-time multimodal classification for astronomy. Direct comparisons with other models from the existing photometric classification literature, even those trained on data from the same survey, proves to be extremely challenging since every group makes different decisions about primary science goals and metric(s) reported. This makes it practically impossible to assess the performance of different methodologies while controlling for other variables. For this reason, we refrain from including direct comparisons between models in the literature for the BTS dataset. Instead, we discuss the design choices and reported performance for some deployed models to help contextualize the \base\ models in the wider ecosystem of photometric classifiers for ZTF.

\texttt{BTSbot} \citep{BTSBot1, Rehemtulla+25_Pretraining} performs binary classification with the goal of finding bright transients ($m\leq18.5$ mag) for automated spectroscopic followup with the SED machine \citep[SEDM;][]{sedm1_Blagorodnova,sedm2_Rigault,sedm3_kim22}. This model does not use full light curves, instead relying on images and metadata alone. On this binary classification task, they report an F1 score of $\sim0.93$, depending on the exact variant used. 

\texttt{Superphot+} \citep{superphot} performs 5-way classification between SN\,Ia, SN\,II, SN\,Ib/c, SN\,IIn, and SLSN-I. Since the model relies on fitting a parametric model to light curves, it requires at least five points of $\text{SNR}\geq3$ per band to produce a good fit, ensuring that there are no events with fewer than ten combined observations. On this pruned dataset, they report a macro F1 score of $0.61 \pm 0.02$ and an accuracy of $0.83 \pm 0.01$.

\texttt{AppleCiDEr} \citep{apple_cider_25} performs 5-way classification between SN\,I, SN\,II, CV, AGN, and TDE. This model is also trained for real-time classification and uses a combination of photometry, metadata, images, and spectra. On this classification task, they report a macro recall score of 0.81. 

Based on this non-exhaustive list, it is clear that different models are developed after making vastly different decisions on the quality, volume, and classes of data that they operate on. As stated above, this makes systematic comparison difficult, underscoring the need for standardized benchmarks in astronomy \citep{starembed}. The models discussed above were chosen since they roughly align with either the depth 1 or depth 2 of our BTS taxonomy (Figure \ref{fig:bts_taxonomy}). Indeed, there are several other models that have also been deployed with more focused science goals \citep[see][for example]{Gomez2020_fleet, sheng24_needle, tdescore}.

Despite the strong performance we report (Section \ref{sec:results}), several major challenges remain for photometric classification. We report noticeably worse performance on underrepresented classes such as SN\,Ibc and SLSN-I, regardless of the modalities used, when compared to classes that are well represented in the training data. There also exists a substantial gap in performance between models built for simulated datasets and models trained on real observations, with our \texttt{ELAsTiCC} models showing much stronger performance across a wider variety of classes. This discrepancy highlights the need for improved methods to develop models that can bridge the gap between the two. Distribution shifts are a major unsolved problem that plagues nearly every field within AI for astronomy \citep{Rehemtulla+26_Nature, DESC-AI-White-Paper}. While transfer learning shows promise for this problem in the context of photometric classification \citep{Gupta25_transfer-learning}, truly unsupervised domain adaptation methods remain largely unexplored in time domain astronomy. 

Furthermore, as our models approach several million parameters \citep[or even a few billion parameters, as with \texttt{AION-1,}][]{Parker25_AION-1}, practical considerations such as throughput and compute use become increasingly important. Policy-based approaches to orchestrate between models of varying size can significantly reduce the total compute used for inference over the duration of a survey. For instance, there is limited utility to using \omni\ over \base\ at late times. Architectural improvements such as mixture-of-experts \citep{Waibel1989_MoE, diz-Leyton25_AstroMoE, apple_cider_25} and training enhancements such as model distillation can potentially help with this objective, while preserving the model's capabilities \citep{Hinton2015DistillingTK}. Reinforcement learning methods also show promise in resolving the full explore tradeoff in sequential decision making settings \citep{Sravan24_RL}.

Finally, as we add high-dimensional modalities such as images to our models, it may be beneficial to use compressed representations instead \citep[such as the ones from \texttt{Minuet},][]{Gagliano25_minuet}. These embeddings could be computed once for every source and then reused as inputs for several models, preventing an explosion in the compute used while retaining the performance enhancements demonstrated in this work. Exploring such approaches remains an important direction for future work.

\section{Conclusion}
\label{sec:conclusion}
In this work, we present new multimodal, hierarchical classifiers for ZTF that can produce reliable, high-level classifications within seconds of the first detection. Our most performant model, \omni, combines light curves, metadata, and images to produce reliable classifications within seconds of the first detection and has been deployed on \texttt{BOOM} \citep{Jegou_du_Laz_25_Boom_Broker}, with classifications made publicly available via \texttt{Fritz} in near real time. To prepare for the next generation of time-domain surveys, we also developed LSST models using the \texttt{ELAsTiCC} dataset achieving performance that is in line with, or better than, other state-of-the-art models. These models also serve as an effective starting point for fine-tuning on real data before deployment on the LSST alert stream. 

Across both real and simulated datasets, we demonstrate that incorporating metadata and image cutouts in addition to light curves substantially improves classification performance at all epochs and across every level of granularity considered, without requiring any additional labeled data. Importantly, these improvements are most apparent at early epochs, when rapid source characterization is critical for prioritizing follow-up observations. As facilities such as LSST \citep{2019Ivezic_LSST}, the Argus Optical Array \citep{Law_2022_argus}, the La Silla Schmidt Southern Survey \citep{miller+25_Ls4}, and the Nancy Grace Roman Space Telescope \citep{Akeson19_roman} increase the volume of transient discoveries, multimodal approaches will become increasingly important for effective real-time triage. 

Beyond improved classification performance, this work introduces practical tools for the broader community. We present an image model that captures salient properties of a transient's local environment without requiring explicit feature engineering or host-galaxy association, and we release our image backbone as a standalone model for integration into other classification pipelines. We also make our hierarchical classification framework publicly available, enabling the rapid development of new hierarchical classifiers from only a dataset and taxonomy specification. 

Finally, by quantifying the trade-offs between classification performance and compute costs across architectures of varying complexity, we provide practical guidance for selecting models that balance scientific performance with latency and computational constraints in real-time deployment scenarios.

\subsection{Code and Model availability} 
\label{sec:code_avail}

The entire \base\ family of models, including the image backbone, are open-weight with the training and evaluation code being available under a modified MIT license. The live version of the code is pip-installable and can be found on our GitHub Repository\footnote{\url{https://github.com/dev-ved30/Oracle}}. All model weights are available via Hugging Face\footnote{\url{https://huggingface.co/collections/vedshah30/oracle}}, allowing for easy integration into future models.

\section{Acknowledgments}
\label{sec:acknowledgements}
Zwicky Transient Facility access for V.G.S., N.R., and A.A.M. was supported by Northwestern University and the Center for Interdisciplinary Exploration and Research in Astrophysics (CIERA).

We gratefully acknowledge the support of the NSF-Simons AI-Institute for the Sky (SkAI) via grants NSF AST-2421845 and Simons Foundation MPS-AI-00010513.

N.R. is supported by a Northwestern University Presidential Fellowship Award. N.R. and A.A.M.~are partially supported by DoE award \#DE-SC0025599. A.A.M.~is also supported by Cottrell Scholar Award \#CS-CSA-2025-059 from the Research Corporation for Science Advancement. N.R.~is also partially supported by NSF grant \# AST-2421845.

This research was supported in part through the computational resources and staff contributions provided for the Quest high performance computing facility at Northwestern University which is jointly supported by the Office of the Provost, the Office for Research, and Northwestern University Information Technology.

Based on observations obtained with the Samuel Oschin Telescope 48-inch and the 60-inch
Telescope at the Palomar Observatory as part of the Zwicky Transient Facility project. ZTF is
supported by the National Science Foundation under Grants No. AST-1440341, AST-2034437,
and currently Award \#2407588. ZTF receives additional funding from the ZTF partnership.
Current members include Caltech, USA; Caltech/IPAC, USA; University of Maryland, USA;
University of California, Berkeley, USA; Cornell University, USA; Drexel University, USA;
University of North Carolina at Chapel Hill, USA; Institute of Science and Technology, Austria;
National Central University, Taiwan, German Center for Astrophysics, Germany, and OKC,
University of Stockholm, Sweden. Operations are conducted by Caltech's Optical Observatory
(COO), Caltech/IPAC, and the University of Washington at Seattle, USA.

\textbf{Software Note: } This work makes use of \texttt{Numpy} \citep{numpy}, \texttt{Astropy} \citep{astropy:2013, astropy:2018, astropy:2022}, \texttt{Scipy} \citep{2020SciPy-NMeth}, \texttt{Matplotlib} \citep{matplotlib}, \texttt{Plotly} \citep{plotly}, \texttt{Pandas} \citep{pandas2}, \texttt{timm} \citep{timm}, \texttt{Umap-learn} \citep{umap}, \texttt{Networkx} \citep{networkX}, \texttt{PyTorch} \citep{pytorch}, \texttt{Weights and Biases} \citep{wandb}, \texttt{Sklearn} \citep{scikit-learn}, and \texttt{Polars}.

\textbf{Data Note: } We want to acknowledge the contributions of the team that created the ELAsTiCC data set: Gautham Narayan, Alex Gagliano, Alex Malz, Catarina Alves, Deep Chatterjee, Emille Ishida, Heather Kelly, John Franklin Crenshaw, Konstantin Malanchev, Laura Salo, Maria Vincenzi, Martine Lokken, Qifeng Cheng, Rahul Biswas, Renée Holžek, Rick Kessler, Robert Knop, Ved Shah Gautam.

\bibliography{sample701}{}
\bibliographystyle{aasjournal}
\clearpage

\appendix
\label{appendix}
\section{Features for BTS classifiers}

\begin{table*}
\centering

\begin{tabular}{l|l}
\multicolumn{2}{c}{\textbf{Contextual Features:}} \\
\toprule
\textbf{Feature} & \textbf{Description} \\
\midrule
\verb|sky| &  Local sky background estimate\\
\verb|sgscore1| &  Star/Galaxy score of the nearest PS1 source\\
\verb|sgscore2| &  Star/Galaxy score of the $2^{\text{nd}}$ nearest PS1 source\\
\verb|distpsnr1| &  Distance (") to the nearest PS1 sources\\
\verb|distpsnr2| &  Distance (") to the $2^{\text{nd}}$ nearest PS1 sources\\
\verb|fwhm| &  Full width half max\\
\verb|diffmaglim| & $5\sigma$ detection threshold (in mag) \\
\verb|ndethist| & Number of previous detections of source \\
\verb|nmtchps| & Number of PS1 cross-matches within 30 arcsec\\
\verb|drb| & Real/Bogus score\\
\verb|ncovhist| & Number of times source on a field and read channel\\
\verb|chinr| &  $\chi$ parameter of nearest source in reference\\
\verb|sharpnr| & sharp parameter of nearest source in reference\\
\verb|scorr| & Peak-pixel signal-to-noise in detection image\\
\verb|sgmag1| & $g$-band magnitude of the nearest PS1 source \\
\verb|srmag1| & $r$-band magnitude of the nearest PS1 source \\
\verb|simag1| &  $i$-band magnitude of the nearest PS1 source \\
\verb|szmag1| & $z$-band magnitude of the nearest PS1 source \\
\verb|sgmag2| & $g$-band magnitude of the $2^{\text{nd}}$ nearest PS1 source \\
\verb|srmag2| & $r$-band magnitude of the $2^{\text{nd}}$ nearest PS1 source \\
\verb|simag2| & $i$-band magnitude of the $2^{\text{nd}}$ nearest PS1 source \\
\verb|szmag2| & $z$-band magnitude of the $2^{\text{nd}}$ nearest PS1 source \\
\verb|W1mag| & $W1$ band magnitude of the nearest WISE source\\
\verb|W2mag| &  $W2$ band magnitude of the nearest WISE source\\
\verb|W3mag| &  $W3$ band magnitude of the nearest WISE source\\
\verb|W4mag| &  $W4$ band magnitude of the nearest WISE source\\
% \verb|W1_minus_W3| & $W1-W3$ color of the nearest WISE source\\
% \verb|W2_minus_W3| & $W2-W3$ color of the nearest WISE source\\
\verb|l| & Galactic longitude of the source\\
\verb|b| & Galactic latitude of the source\\
\bottomrule
\end{tabular}

\bigskip

\begin{tabular}{l|l}
\multicolumn{2}{c}{\textbf{Time-Dependent Features:}} \\
\toprule
\textbf{Feature} & \textbf{Description}\\
\midrule
\verb|magpsf| & Magnitude of the detection \\
\verb|sigmapsf| & $1\sigma$ uncertaintiy on the magnitude\\
\verb|days| & Days since first detection\\
\verb|filter| & Mean wavelength of the passband \\
\verb|photflag| & Flags for detections and non-detections\\
\bottomrule
\end{tabular}

\caption{Description of features for the metadata backbone (top) and the time series backbone (bottom) used to train the BTS models presented in this work.}
\label{table:bts_featues}
\end{table*}

\clearpage
\section{Features for elasticc classifiers}

\begin{table*}
\centering

\begin{tabular}{l|l}
\multicolumn{2}{c}{\textbf{Time-Independent Features:}} \\
\toprule
\textbf{Feature} & \textbf{Description} \\
\midrule
\verb|MWEBV| & Milky way extinction\\
\verb|MWEBV_ERR| & Error in milky way extinction\\
\verb|REDSHIFT_HELIO| &  Best heliocentric redshift. z-Spec if available; else z-Phot\\
\verb|REDSHIFT_HELIO_ERR| &  Error in best heliocentric redshift\\
\verb|HOSTGAL_PHOTOZ| &  z-Phot for the Host Galaxy if available\\
\verb|HOSTGAL_PHOTOZ_ERR| &   Error in z-Phot for the Host Galaxy\\
\verb|HOSTGAL_SPECZ| &  z-Spec for the Host Galaxy if available\\
\verb|HOSTGAL_SPECZ_ERR| &  Error in z-Spec for the Host Galaxy\\
\verb|HOSTGAL_RA| &  RA for the Host Galaxy\\
\verb|HOSTGAL_DEC| &  Dec for the Host Galaxy\\
\verb|HOSTGAL_SNSEP| & Transient-host separation, in arcsec\\
\verb|HOSTGAL_ELLIPTICITY| & Ellipticity of the Host Galaxy\\
\verb|HOSTGAL_MAG_[u,g,r,i,z,Y]| &  [u,g,r,i,z,Y] - band magnitudes for the Host Galaxy\\
\bottomrule
\end{tabular}

\bigskip

\begin{tabular}{l|l}
\multicolumn{2}{c}{\textbf{Time-Dependent Features:}} \\
\toprule
\textbf{Feature} & \textbf{Description}\\
\midrule
\verb|FLUXCAL| & The calibrated flux value from SNANA \\
\verb|FLUXCAL_ERR| &  Uncertainty on FLUXCAL from SNANA \\
\verb|TIME| & Days since first observation\\
\verb|BAND| & Mean wavelength of the passband in $\mu m$\\
\verb|PHOTFLAG| & Flags for detections (1) and non-detections (0)\\
\bottomrule
\end{tabular}

\caption{Description of metadata (top) and time series features (bottom) used to train the \texttt{ELAsTiCC} models presented in this work.}
\label{table:elasticc_featues}
\end{table*}

\clearpage
\section{Hyperparameter sweeps for BTS Models.}

\begin{table*}[]
    \centering
    \begin{tabular}{c|cccc}
    \toprule 
     \textbf{} & Image Backbone & \lite & \base & \omni \\
    \midrule
    \texttt{lr} & [5e-7, \textbf{4e-5}, 5e-5] & [5e-5, \textbf{2.5e-4}, 5e-3] & [5e-5, \textbf{4.5e-5}, 5e-3] & [5e-7, \textbf{1e-4}, 5e-4]\\
    \texttt{batch size} & \{32, 64, \textbf{128}\} & \{32, \textbf{64}, 128, 256\} & \{32, \textbf{64}, 128, 256\} & \{\textbf{32}, 64, 128\} \\
    $\alpha$ & [0, \textbf{0.49}, 0.5] & [0, \textbf{0.34}, 0.5] & [0, \textbf{0.31}, 0.5] & [0, \textbf{0.07}, 0.5]\\
    warm-up & -- & -- & -- & [None, 50, \textbf{100}]\\
    % \midrule
    % LSST  & \texttt{lr} & -- & [] & [] & --\\
    %   & \texttt{batch size} & -- & \{\} & \{\} & -- \\
    %   & $\alpha$ & -- & [] & [] & --\\
    %   % & $\gamma$ & -- & -- & -- & \\
    \bottomrule
    \end{tabular}
    \caption{Hyperparameters explored while training the BTS models present in this work. The best hyperparameters, which were used to train the final models, are highlighted in bold.}
    \label{tab:hp_tuning}
\end{table*}

\clearpage
\section{Performance on the ELAsTiCC Dataset}

\begin{table*}[]
    \centering
    \begin{tabular}{l|cccc|cccc}
        \toprule
        \toprule
         \multicolumn{1}{c|}{} & \multicolumn{4}{c|}{\lite} & \multicolumn{4}{c}{\base} \\
         & $F1_{1}$ & $F1_{8}$ & $F1_{64}$ & $F1_{1024}$ & $F1_{1}$ & $F1_{8}$ & $F1_{64}$ & $F1_{1024}$ \\
        \midrule
        \midrule
        \multicolumn{9}{c}{\textbf{Depth 1}} \\
        \midrule
        Transient & 0.98±0.00 & 0.98±0.00 & \textbf{1.00±0.00} & \textbf{1.00±0.00} & \textbf{0.99±0.00} & \textbf{0.99±0.00} &\textbf{1.00±0.00} & \textbf{1.00±0.00} \\
        Variable & 0.94±0.00 & 0.95±0.00 & 0.99±0.00 & \textbf{1.00±0.00} & \textbf{0.96±0.00} & \textbf{0.97±0.00} & \textbf{1.00±0.00} & \textbf{1.00±0.00} \\
        \midrule
        % accuracy & 0.96±0.00 & 0.97±0.00 & 1.00±0.00 & 1.00±0.00 & 0.98±0.00 & 0.98±0.00 & 1.00±0.00 & 1.00±0.00 \\
        \textbf{macro} & 0.96±0.00 & 0.97±0.00 & \textbf{1.00±0.00} & \textbf{1.00±0.00} & \textbf{0.97±0.00} & \textbf{0.98±0.00} & \textbf{1.00±0.00} & \textbf{1.00±0.00} \\
        % weighted avg & 0.97±0.00 & 0.97±0.00 & 1.00±0.00 & 1.00±0.00 & 0.98±0.00 & 0.98±0.00 & 1.00±0.00 & 1.00±0.00 \\
        \midrule
        \midrule
        \multicolumn{9}{c}{\textbf{Depth 2}} \\
        \midrule
        AGN & 0.74±0.01 & 0.81±0.01 & 0.95±0.00 & 0.99±0.00 & \textbf{0.93±0.00} & \textbf{0.95±0.00} & \textbf{0.99±0.00} & \textbf{1.00±0.00} \\
        Fast & 0.83±0.01 & 0.90±0.00 & 0.97±0.00 & 0.98±0.00 & \textbf{0.90±0.00} & \textbf{0.94±0.00} & \textbf{0.99±0.00} & \textbf{0.99±0.00} \\
        Long & 0.72±0.00 & 0.76±0.00 & 0.88±0.00 & 0.90±0.00 & \textbf{0.80±0.00} & \textbf{0.83±0.00} & \textbf{0.91±0.00} & \textbf{0.93±0.00} \\
        Periodic & 0.90±0.00 & 0.93±0.00 & 0.99±0.00 & \textbf{1.00±0.00} & \textbf{0.97±0.00} & \textbf{0.98±0.00} & \textbf{1.00±0.00} & \textbf{1.00±0.00} \\
        SN & 0.69±0.00 & 0.75±0.00 & 0.88±0.00 & 0.90±0.00 & \textbf{0.80±0.00} & \textbf{0.84±0.00} & \textbf{0.92±0.00} & \textbf{0.93±0.00} \\
        \midrule
        % accuracy & 0.77±0.00 & 0.82±0.00 & 0.93±0.00 & 0.94±0.00 & 0.86±0.00 & 0.89±0.00 & 0.95±0.00 & 0.96±0.00 \\
        \textbf{macro} & 0.78±0.00 & 0.83±0.00 & 0.94±0.00 & 0.95±0.00 & \textbf{0.88±0.00} & \textbf{0.91±0.00} & \textbf{0.96±0.00} & \textbf{0.97±0.00} \\
        % weighted avg & 0.77±0.00 & 0.83±0.00 & 0.93±0.00 & 0.94±0.00 & 0.86±0.00 & 0.89±0.00 & 0.95±0.00 & 0.96±0.00 \\
        \midrule
        \midrule
        \multicolumn{9}{c}{\textbf{Depth Leaf}} \\
        \midrule
        AGN & 0.63±0.01 & 0.73±0.01 & 0.95±0.00 & 0.99±0.00 & \textbf{0.92±0.00} & \textbf{0.94±0.00} & \textbf{0.99±0.00} & \textbf{1.00±0.00} \\
        CART & 0.24±0.01 & 0.31±0.01 & 0.48±0.02 & 0.53±0.02 & \textbf{0.38±0.01} & \textbf{0.44±0.01} & \textbf{0.61±0.01} & \textbf{0.65±0.01} \\
        Cepheid & 0.79±0.01 & 0.84±0.01 & 0.98±0.00 & \textbf{0.99±0.00} & \textbf{0.86±0.01} & \textbf{0.89±0.01} & \textbf{0.99±0.00} & \textbf{0.99±0.00} \\
        Delta Scuti & 0.60±0.01 & 0.70±0.01 & 0.96±0.00 & \textbf{0.99±0.00} & \textbf{0.66±0.01} & \textbf{0.74±0.01} & \textbf{0.97±0.00} & \textbf{0.99±0.00} \\
        Dwarf Novae & 0.89±0.00 & 0.91±0.00 & 0.96±0.00 & 0.96±0.00 & \textbf{0.93±0.00} & \textbf{0.95±0.00} & \textbf{0.97±0.00} & \textbf{0.97±0.00} \\
        EB & 0.81±0.00 & 0.86±0.01 & \textbf{0.98±0.00} & \textbf{0.99±0.00} & \textbf{0.86±0.00} & \textbf{0.90±0.00} & \textbf{0.98±0.00} & \textbf{0.99±0.00} \\
        ILOT & 0.45±0.00 & 0.49±0.01 & 0.71±0.01 & 0.84±0.00 & \textbf{0.60±0.01} & \textbf{0.62±0.00} & \textbf{0.81±0.01} & \textbf{0.88±0.01} \\
        KN & 0.59±0.02 & 0.78±0.00 & 0.91±0.00 & 0.93±0.00 & \textbf{0.79±0.00} & \textbf{0.90±0.00} & \textbf{0.97±0.00} & \textbf{0.98±0.00} \\
        M-dwarf Flare & 0.89±0.00 & 0.93±0.00 & 0.94±0.00 & 0.95±0.00 & \textbf{0.93±0.00} & \textbf{0.96±0.00} & \textbf{0.97±0.00} & \textbf{0.97±0.00} \\
        PISN & 0.49±0.01 & 0.53±0.01 & 0.79±0.01 & 0.90±0.01 & \textbf{0.73±0.00} & \textbf{0.75±0.01} & \textbf{0.88±0.00} & \textbf{0.93±0.00} \\
        RR Lyrae & 0.59±0.01 & 0.67±0.01 & 0.96±0.00 & \textbf{0.99±0.00} & \textbf{0.65±0.01} & \textbf{0.72±0.01} & \textbf{0.97±0.00} & \textbf{0.99±0.00} \\
        SLSN & 0.47±0.00 & 0.52±0.01 & 0.70±0.01 & 0.80±0.01 & \textbf{0.69±0.00} & \textbf{0.73±0.01} & \textbf{0.84±0.01} & \textbf{0.88±0.00} \\
        SNI91bg & 0.48±0.00 & 0.59±0.01 & 0.78±0.01 & 0.80±0.01 &\textbf{ 0.67±0.02} & \textbf{0.74±0.01} & \textbf{0.86±0.01} & \textbf{0.87±0.01} \\
        SNII & 0.22±0.01 & 0.30±0.01 & 0.53±0.00 & 0.58±0.01 & \textbf{0.32±0.01} & \textbf{0.39±0.01} & \textbf{0.60±0.01} & \textbf{0.65±0.01} \\
        SNIa & 0.34±0.00 & 0.42±0.01 & 0.62±0.01 & 0.66±0.01 & \textbf{0.51±0.01} & \textbf{0.58±0.01} & \textbf{0.72±0.01} & \textbf{0.75±0.00} \\
        SNIax & 0.29±0.01 & 0.37±0.01 & 0.53±0.00 & 0.57±0.01 & \textbf{0.38±0.01} & \textbf{0.46±0.01} & \textbf{0.60±0.01} & \textbf{0.64±0.01} \\
        SNIb/c & 0.13±0.00 & 0.20±0.01 & 0.45±0.01 & 0.48±0.01 & \textbf{0.34±0.02} & \textbf{0.42±0.01} & \textbf{0.58±0.01} & \textbf{0.60±0.00} \\
        TDE & 0.55±0.01 & 0.65±0.01 & 0.89±0.00 & 0.91±0.00 & \textbf{0.61±0.00} & \textbf{0.71±0.00} & \textbf{0.90±0.00} & \textbf{0.93±0.01} \\
        uLens & 0.75±0.00 & 0.80±0.01 & 0.93±0.01 & 0.95±0.00 & \textbf{0.82±0.01} & \textbf{0.87±0.01} & \textbf{0.96±0.00} & \textbf{0.96±0.00} \\
        \midrule
        % accuracy & 0.54±0.00 & 0.62±0.00 & 0.79±0.00\textbf{} & 0.84±0.00 & 0.67±0.00 & 0.72±0.00 & 0.85±0.00 & 0.88±0.00 \\
        \textbf{macro} & 0.54±0.00 & 0.61±0.00 & 0.79±0.00 & 0.83±0.00 & \textbf{0.67±0.00} & \textbf{0.72±0.00} & \textbf{0.85±0.00} & \textbf{0.88±0.00} \\
        % weighted avg & 0.53±0.00 & 0.61±0.00 & 0.79±0.00 & 0.83±0.00 & 0.66±0.00 & 0.72±0.00 & 0.85±0.00 & 0.87±0.00 \\
        \bottomrule
        \bottomrule
    \end{tabular}
    \caption{Per-class F1 and macro F1 for the \lite\ (light curve only) and \base\ (light curve + metadata) models, across all levels of the \texttt{ELAsTiCC} taxonomy, at various phases of light curve evolution. The best performance (within $1\sigma$ uncertainties) on each metric is highlighted in bold.}
    \label{tab:f1_elasticc}
\end{table*}

\begin{figure*}
    \centering
    \includegraphics[width=\linewidth]{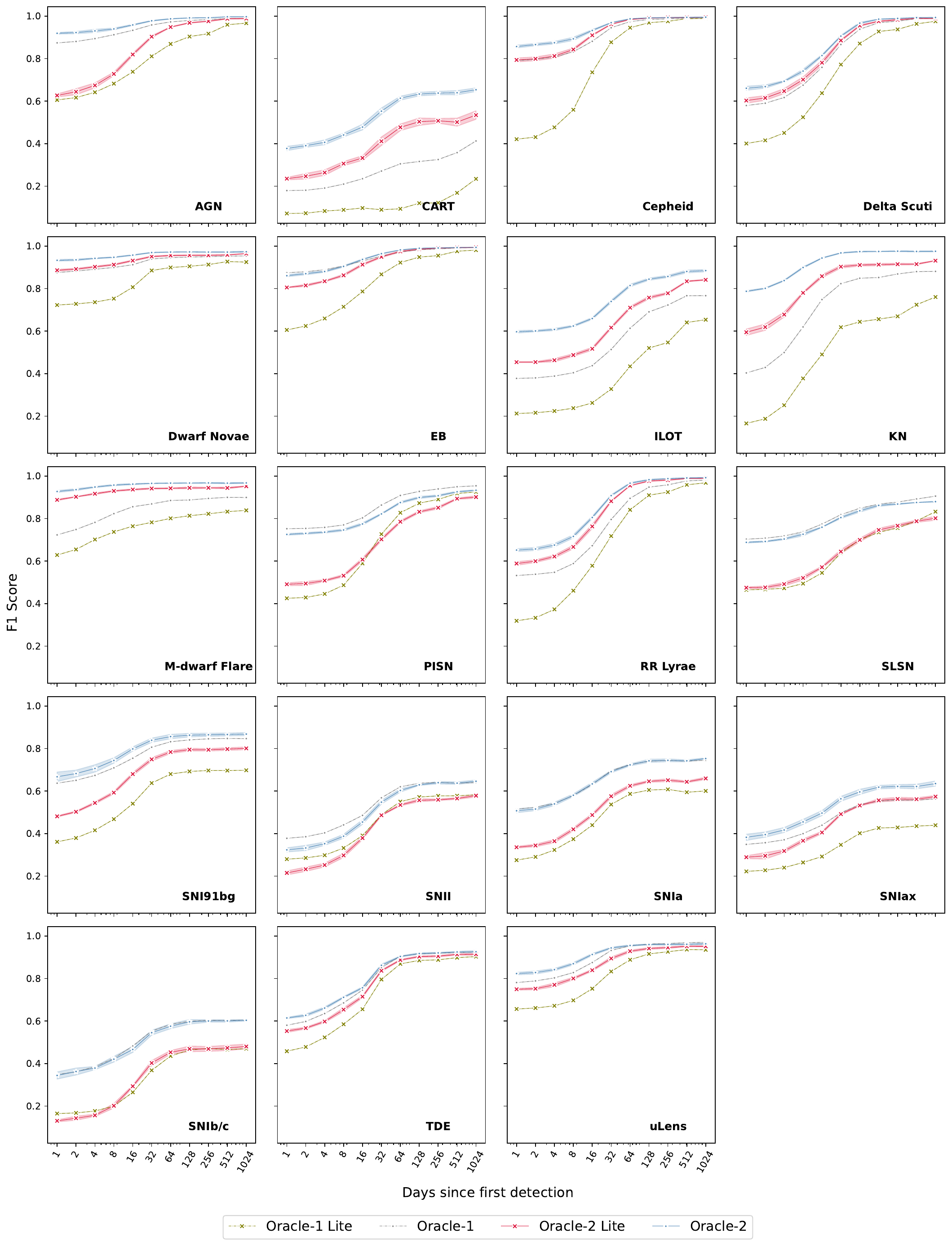}
    \caption{Time evolution of the class F1 scores for the \lite\ (light curve only) and \base\ (light curve + metadata) models for all 19 leaf classes in the \texttt{ELAsTiCC} dataset.}
    \label{fig:oracle2_elasticc_cf_per_class_f1}
\end{figure*}

\end{document}